\documentclass{jfm}

\usepackage{graphicx}
\usepackage{newtxtext}
\usepackage{newtxmath}
\usepackage{natbib}

\usepackage{color} 
\usepackage{hyperref}
\usepackage{amsmath}
\usepackage{subfigure}

\let\Gamma\varGamma
\let\Theta\varTheta
\let\Lambda\varLambda
\let\Xi\varXi
\let\Pi\varPi
\let\Sigma\varSigma
\let\Upsilon\varUpsilon
\let\Phi\varPhi
\let\Psi\varPsi
\let\Omega\varOmega

\newcommand{\vect}[1]{\boldsymbol{#1}}
\hypersetup{
    colorlinks = true,
    urlcolor   = blue,
    citecolor  = blue,
}
\newcommand{\RomanNumeralCaps}[1]
\graphicspath{{./figs/}}

\title{A quasi-dynamic one-equation model with joint constraints of kinetic energy and helicity fluxes for large eddy simulation of rotating turbulence}

\author{Depei Song\aff{1,2}, Changping Yu\aff{1}\corresp{\email{cpyu@imech.ac.cn}}, Zheng Yan\aff{3} and Xinliang Li\aff{1,2}}

\affiliation{\aff{1}LHD, Institute of Mechanics, Chinese Academy of Sciences, Beijing 100190, PR China
\aff{2}School of Engineering Science, University of Chinese Academy of Sciences, Beijing 100049, PR China
\aff{3}Institute of Applied Physics and Computational Mathematics, Beijing 100094, PR China}

\begin{document}
\maketitle

\begin{abstract}
	For settling the problem with rotating turbulence modelling, a quasi-dynamic one-equation subgrid-scale (SGS) model is proposed in this paper. Considering the key role of the joint cascade of kinetic energy and helicity in rotating turbulence, the new SGS model is constrained by the fluxes of kinetic energy and helicity. Specifically, the new theory of dual channels of helicity flux is taken into account. The modelling of the unclosed quantities is achieved by adopting a quasi-dynamic process that eliminates the need for test filtering compared to the classic dynamic process, and the model coefficients are dynamically obtained through the SGS kinetic energy transport equation and considering the joint constraints of kinetic energy and helicity fluxes. As a result, the model demonstrates a high correlation with DNS data in \textit{a priori} tests. We refer to this new model as the quasi-dynamic joint-constraint model (QCM), which is introduced for both incompressible and compressible flows. To assess the effectiveness of the QCM, numerical experiments are conducted for three typical cases: incompressible streamwise rotating channel flow, transonic streamwise rotating annular pipe flow, and hypersonic transition flow at Mach 6 over a rotating cone. The results suggest that the QCM has the potential to significantly improve the prediction of rotational flows that are strongly influenced by helicity. Additionally, the new model demonstrates excellent capability in handling the transition process.
\end{abstract}


{\bf MSC Codes }  {\it(Optional)} Please enter your MSC Codes here

\section{Introduction}
Rotating turbulence is of critical importance in wide fields of scientific, engineering, and product design applications \citep{childs2010rotating}, including the design of pumps, gas turbines, and the study of atmospheric and oceanic flows \citep{wu2004effects, weller2006dns}. In the field of hypersonic high-temperature gas dynamics, the design and development of rotating detonation engines also require precise simulation of rotating turbulence \citep{Lu2014RotatingDW}. Furthermore, the investigation of rotating flow is important to the study of turbulence. Turbulent flows subjected to system rotations have been extensively studied in the literature. \cite{johnston1972effects} conducted experiments on fully developed turbulent flow in a rotating channel and observed three stability-related phenomena, which indicates the complex nature of this kind of flow. In this paper, our objective is to establish an accurate subgrid-scale (SGS) model for large-eddy simulation (LES) of rotating turbulence.

Large-eddy simulation is a technique to resolve large-scale motions while modelling the dynamic behaviour of small-scale motions that cannot be adequately resolved on a computational mesh \citep{jameson2022computational}. This method was pioneered for meteorological flows by \citet{smagorinsky1963}. The Smagorinksy model (SM) is a subgrid-scale stress model with eddy-viscosity formulation and \citet{lilly1967representation} derived the Smagorinsky coefficient based on the Kolmogorov theory \citep{kolmogorov1941local}. Great success has been achieved in simulations of decaying homogeneous isotropic turbulence \citep{versteeg2007introduction}; however, this model is not suitable for transitional or wall-bounded flows due to the modelled SGS stress does not vanish in near-wall regions or laminar regions \citep{blazek2015computational}. After the Smagorinsky model, many other eddy-viscosity models have been developed. The dynamic Smagorinsky model (DSM) developed by \citet{germano1991} addresses the issues of excessive dissipation and the variation of model coefficients in different flows in the SM model by introducing test filtering. This helps to mitigate excessive dissipation and ensures that the model coefficients are adaptive to different flow conditions.
The dynamic Smagorinsky model also has its limitations, to ensure numerical stability, special treatment such as clipping or spatial averaging is often needed \citep{garnier2009large}. Therefore, the improved dynamic SGS models \citep{ghosal1995,carati1995} were proposed. Ducros \citep{ducros1998wall} proposed the wall-adapting local eddy-viscosity (WALE) model, which can have the correct asymptotic behaviour near the wall. Therefore, the damping procedure is not necessary for SGS viscosity in the near-wall region. The Vreman model \citep{vreman2004eddy} is another eddy-viscosity model and it has a simple form and better near-wall behaviour. Moreover, the eddy-viscosity model and its variants are based on the hypothesis that the subgrid-scale (SGS) stress aligns with the large-scale strain rates. However, a \textit{priori} analysis using direct numerical simulation (DNS) data has revealed that this assumption is invalid for most canonical flows, as the SGS stresses are poorly correlated with the large-scale strain rates \citep{bardina1980improved,meneveau2000scale}. To overcome the drawbacks of peer dissipation and low correlation with the real flow in eddy-viscosity models, the non-linear model which does not depend on the Boussinesq hypothesis is proposed. The non-linear model that has tensorial eddy viscosity usually shows a higher correlation with the local subgrid stress. However, a number of non-linear models have observed issues in providing adequate dissipation of turbulent kinetic energy \citep{clark1979evaluation,stolz1999}.

Subsequently, the deconvolution methods developed by \citet{stolz1999} can be applied to derive the similarity model \citep{bardina1980improved} and the Clark model \citep{clark1979evaluation}. The deconvolution models show many realistic features when compared with real SGS stress fields in \textit{a priori} tests, but tend to produce non-physical behaviour when implemented in simulations. To overcome the difficulty of the deconvolution model, \citet{zang1993} and \citet{armenio2000} proposed a mixed model that adds additional dissipative mechanisms to the original model. \citet{domaradzki2002} showed that perfect deconvolution is equivalent to under-resolved DNS. It is worth mentioning that the aforementioned models are purely algorithmic, as they depend solely on the definitions of the filter and the SGS stress tensor, without demanding any physical modelling.

% However, additional dissipation makes it impossible to recover the realistic performance of perfect deconvolution and the dissipation should be anisotropic, which still brings problem for the modelling of dissipative mechanism.

Regarding rotating turbulence, several SGS models have also been proposed. \citet{Piomelli1995} suggested a localized dynamic model and validated the accuracy of the model in the case of streamwise-rotating channel flow. \citet{lamballais1998spectral} proposed a spectral-dynamic model based on the Eddy Damped Quasi-Normal Markovan (EDQNM) theory, the new model shows an improved prediction of near-wall behaviour in the case of streamwise-rotating channel flow. \citet{yang2004large} conducted the large-eddy simulation of decaying rotating turbulence using the truncated Navier-Stokes method but limited it to a low Reynolds number. \citet{kobayashi2005subgrid} proposed an SGS model based on coherent structures that aims for rotating flows and got improved performance compared with dynamic Smagorinksy models, but this approach is difficult to generalize to compressible flows. \citet{li2006subgrid} showed that the Smagorinksy model tends to underpredict helicity dissipation rate in rotating turbulence, and they proposed a dynamic model considering both energy and helicity dissipation balance. \citet{baerenzung2008spectral} proposed an LES model based on the evaluation of energy and helicity cascade, but this model is computationally expensive and is hard to be applied at complex geometry.
% Compare to classical fluid dynamics, there are relatively fewer developed SGS models for rotating turbulence. In this paper, we attempt to establish a novel LES model specifically designed for rotational flow. Our primary focus in developing an effective LES model is to consider the phenomenon of turbulence cascade within the turbulent flows.

% The Energy cascade process is vitally important for LES modelling, the famous $-5/3$ exponent of the turbulent energy spectrum has been validated in all kinds of turbulent flows thus a variety of LES models have been derived for modelling the forward energy cascade process, the Smagorinksy model and models based on subgrid-scale kinetic energy equation can be put in this category \citep{Yoshizawa1982ASS, yoshizawa1985}. However, the K41 theory doesn't consider the intermittency phenomenon in anisotropic flows, and it also neglected the possible backscatter of energy. \citet{chasnov1991} developed a spectral model based on the Eddy Damped Quasi-Normal Markovan (EDQNM) model that modelled the backward energy cascade process, which confirmed that negative subgrid viscosity can be used to take into account the backward cascade effects that are dominant for very small wavenumbers. The presence of negative eddy viscosity also explained why models base on Boussinesq eddy viscosity assumption like the Smagorinksy model are often too dissipative in practical application.

In rotating turbulent flows, the joint cascades of kinetic energy and helicity always occur together, and they are key physical processes in rotating turbulence \citep{andre1977influence}. 
% The energy cascade, as in canonical flows, is thoroughly studied. However, the study of helicity cascade is rarely seen in the literature. To understand the nature of both intermittency and backward energy cascade, effects of helicity should be further verified.
Helicity is one of the two quadratic inviscid invariants of Navier-Stokes equation \citet{moffatt2014}, and it can be defined as the product of velocity and vorticity, i.e. $h = \vect{u}\cdot \vect{\omega}$.
% Defined as the integration over all space of  the product of velocity and vorticity, i.e. $H(t) = \int \vect{u}\cdot \vect{\omega} dV$, 
Helicity is a measure of the degree of knottedness and/or linkage of the vortex lines of the flow \citep{moffatt1969degree}.
Like energy cascade in the classic Kolmogorov theory, helicity cascade also plays an important role in the evolution of turbulence by affecting the energy cascade process, either promoting the inversion of the energy cascade or impeding the forward energy cascade \citep{chen2003,biferale2012,plunian2020}. It has confirmed that there is a kinematic connection between energy cascade and helicity and this connection plays an essential role in large-scale intermittency \citep{baj2022}. Through theoretical and numerical analysis, the joint cascade of energy and helicity at high Reynolds number has been revealed, and the role of helicity is even important in rotating helical turbulence as it dominates the process of energy cascade \citep{mininni2009helicity,mininni2010rotating}. Moreover, in stream-wise rotation channel flows, the distribution of the mean spanwise secondary flow and the near-wall Taylor-G{\"o}tler vortices are closely related to the high helicity \citep{yu2022}.
% lead to a mean secondary flow in the spanwise thus indicates the existence of high helicity. For this reason, it is vitally important to consider the possible effects of helicity on the energy cascade in LES modelling of rotating helical turbulent flows. 
% For LES modelling, the kinetic energy flux which represents transfer of energy to small scales is a core physical quantity for the turbulent kinetic energy cascade \citep{moser2021statistical}. Similarly, the helicity flux is a representative quantity of helicty cascade and key statistical quantity in LES for rotating turbulence. 
Recently, \citet{yan2020} has proved that there is a dual channel mechanism in the helicity cascade process, and the new theory gives a proper description of helicity flux in different types of turbulent flows. Furthermore, the dual channel helicity fluxes have apparently different properties, for example, the two channels are dominated by vortex extension and vortex distortion processes respectively \citep{yu2022}. Distinguished from traditional single-channel theory, the new theory of dual channels of helicity flux will be
more suitable for highly anisotropic rotating turbulent flows.
% dual channels of helicity flux exists in rotating turbulence and the two channels are dominated by vortex extension and vortex distortion processes respectively. It has been found that in anisotropic flows, the second channel may cause the inverse energy cascades, which is vitally important for modelling the sub-grid scale mechanisms in rotating flows. Another work of \citet{baj2022} has confirmed that there is a kinematic connection between energy cascade and helicity and this connection plays an essential role in large-scale intermittency. 
For LES modelling, the kinetic energy flux (KEF) is a core physical quantity for the turbulent kinetic energy cascade \citep{moser2021statistical}. Similarly, helicity flux and kinetic energy flux are also the core quantities of the joint cascade of kinetic energy and helicity in LES of rotating turbulence. As almost all the previous work of large-eddy simulation on rotating flows did not take into account the dual channel effect of helicity, this may hinder accurate representation of the energy and helicity cascade process.

% Based on the theory of dual channels of helicity flux, this study proposes a new quasi-dynamic SGS helicity flux dual channel model (QCM). Distinguished from traditional single-channel theory, the new model based on the theory of dual channels of helicity flux will be more suitable for highly anisotropic rotating turbulent flows. 
To accurately simulate rotating turbulence using large eddy simulation, we introduce physical constraints and numerical stability constraints as mentioned in \citet{sagaut2005large}. The physical constant, which means the model must be consistent from the viewpoint of the phenomenon being modelled, is crucially important for the accuracy of the LES model. To achieve this, the joint constraints kinetic and helicity fluxes based on the new theory of dual channels of helicity flux will be introduced to construct the new LES model. We start with the modelling of the energy cascade and helicity cascade process, which are the two representative phenomena in turbulence. The numerical stability constraint means the model must not destabilize the numerical simulation and be insensitive to discretization. To maintain numerical stability, we adopt the quasi-dynamic process proposed in our previous work \citep{qi2022a}. By using precisely resolved SGS turbulent kinetic energy, turbulent kinetic energy and helicity fluxes to constrain the Smagorinsky model, more accurate model coefficients can be obtained, ensuring numerical stability without sacrificing accuracy. It is expected that this model can accurately depict kinetic energy and helicity cascades in rotating flows.

The paper is organized as follows: the LES modelling for incompressible flows is introduced in \S{2}. The verification of the model in the streamwise rotating channel is supplied in \S{2.4}, where detailed $priori$ and $posteriori$ tests are presented. The derivation of the new model in compressible flows is supplied in \S{3}. In \S{3.4} and \S{3.5}, we test the new model in two representative cases: the transonic flow in a rotating annular pipe and the hypersonic flow over a rotating cone. Finally, the conclusions are provided in \S{4}.

\section{LES modelling for incompressible flows}
\subsection{Governing equations of LES}
For the incompressible turbulent flows, the filtered Navier-Stokes equations are taken as follows:
\begin{subeqnarray}
	\frac{\partial \bar{u}_i}{\partial x_i} &=& 0 , \slabel{eq:continuity} \\[3pt]
	\frac{\partial \bar{u}_i}{\partial t} + \bar{u}_j\frac{\partial \bar{u}_i}{\partial x_j} = -\frac{1}{\rho}\frac{\partial \bar{p}}{\partial x_i} &+& \nu\frac{\partial^2 \bar{u}_i}{\partial x_j^2} + \bar{f}_i -\frac{\partial \tau_{ij}}{\partial x_j}. \slabel{eq:momentum}
\end{subeqnarray}
where $\left(\bar{\cdot}\right)$ denotes spatial filtering at scale $\Delta$, $\bar{f}_i$ is the filtered forcing, and $\tau_{ij}=\overline{u_iu_j}-\bar{u}_i\bar{u}_j$ is the SGS stress tensor that needs to be modelled. The isotropic part of the SGS stress is often absorbed into pressure and eddy viscosity based SGS models based on the Boussinesq hypothesis take the form
\begin{equation}
	\tau_{ij}-\frac{1}{3}\tau_{kk}\delta_{ij} = -2\nu_{sgs}\bar{S}_{ij}, \quad \bar{S}_{ij} = \frac{1}{2}\left(\frac{\partial \bar{u}_i}{\partial x_j} + \frac{\partial \bar{u}_j}{\partial x_i}\right),
\end{equation}
where $\delta_{ij}$ is the Kronecker-delta function.

In the absence of system rotation, the filtered forcing term in momentum equation~(\ref{eq:momentum}) contains the Coriolis force term $-2\vect{\Omega}\times \vect{\bar{u}}$, where $\bar{\vect{u}}$ is the filtered velocity and $\vect{\Omega}$ is the rotating vector of the system. The pressure $\bar{p}$ is the total pressure containing both the static pressure and the centrifugal force.

\subsection{Dual channels of helicity flux and the SGS kinetic energy equation}
% By taking curl of the momentum equation \ref{eq:momentum}, we can obtain the filtered vorticity equation:
% \begin{equation}
% 	\frac{\partial \bar{\omega}_i}{\partial t}+\bar{u}_j \frac{\partial \bar{\omega}_i}{\partial x_j}=\bar{\omega}_j \frac{\partial \bar{u}_i}{\partial x_j}+v \frac{\partial^2 \bar{\omega}_i}{\partial x^2_j}-\frac{\partial \gamma_{i j}}{\partial x_j},
% \end{equation}
% where $\bar{\vect{\omega}}$ is the filtered vorticity and $\gamma_{ij} = \left(\overline{\omega_i u_j}-\bar{\omega}_i\bar{u}_j\right)-\left(\overline{\omega_j u_i}-\bar{\omega}_j\bar{u}_i\right)$ is the SGS vortex stretching stress. From this equation we can derive the transport equations of large scale helicity $h_{\Delta} = \bar{\vect{u}}\cdot \bar{\vect{\omega}}$ \citep{yan2020}:
% \begin{equation}
% 	\partial_t h_{\Delta} + \nabla \cdot \vect{Q} = -\Pi_{\Delta}^{H1}-\Pi_{\Delta}^{H2}-4\nu\bar{\vect{S}}:\bar{\vect{R}}, \label{eq:dual}
% \end{equation}
% where $\bar{\vect{R}} = \frac{1}{2}\left(\nabla \bar{\vect{\omega}}+ \left(\nabla \bar{\vect{\omega}}\right)^T\right)$ and $\vect{Q}$ is the spatial transport of large-scale helicity, which is defined as 
% \begin{equation}
% 	\boldsymbol{Q}=h_{\Delta} \bar{\boldsymbol{u}}+\bar{\boldsymbol{\omega}} \cdot \boldsymbol{\tau}+\bar{\boldsymbol{u}} \cdot \boldsymbol{\gamma}+\frac{\bar{p}}{\rho} \bar{\boldsymbol{\omega}}-\frac{1}{2}|\bar{\boldsymbol{u}}|^2 \bar{\boldsymbol{\omega}}-2 v(\bar{\boldsymbol{u}} \cdot \bar{\boldsymbol{R}}+\bar{\boldsymbol{\omega}} \cdot \bar{\boldsymbol{S}}) ,
% \end{equation}

According to \citet{yan2020}, the first and second channel of helicity flux $\Pi_{\Delta}^{H1}, \Pi_{\Delta}^{H2}$ are expressed as
\begin{equation}
	\Pi_{\Delta}^{H 1}=-\tau_{ij}\bar{R}_{ij}, \quad \Pi_{\Delta}^{H 2}=-\gamma_{ij}\bar{\Omega}_{ij}.
\end{equation}
Here, $\bar{\Omega}_{ij} = \frac{1}{2}\left(\partial\bar{u}_i/\partial x_j - \partial\bar{u}_j/\partial x_i \right)$, $\bar{R}_{ij} = \frac{1}{2}\left(\partial \bar{\omega}_i/\partial x_j + \partial \bar{\omega}_j/\partial x_i \right)$ is the large-scale rotation rate and the large-scale symmetric vorticity gradient, respectively. $\gamma_{i j}=\left(\overline{\omega_i u_j}-\bar{\omega}_i \bar{u}_j\right)-\left(\overline{\omega_j u_i}-\bar{\omega}_j \bar{u}_i\right)$ is the subgrid-scale vortex stretching stress. 

The kinetic energy flux which represents energy transferred to small scales, is defined as
\begin{equation}
	\Pi_{\Delta}^E = \tau_{ij}\bar{S}_{ij},
\end{equation}
where $\bar{S}_{ij} = \frac{1}{2}\left(\partial\bar{u}_i/\partial x_j + \partial\bar{u}_j/\partial x_i \right)$ is the large-scale strain rate tensor.

Following the method of quasi-dynamic procedure, we derive SGS kinetic energy equation for both incompressible and compressible flows. For incompressible flow, the evolution equation of filtered sub-grid kinetic energy $k_{sgs} = \overline{u_k^{\prime}u_k^{\prime}}/2$ can be expressed as:
\begin{equation}
	\begin{split}
		\frac{\partial k_{sgs}}{\partial t} = &-\frac{\partial}{\partial x_j}(k_{sgs}\bar{u}_j)-\frac{1}{2}\frac{\partial}{\partial x_j}(\overline{u_iu_iu_j}-\overline{u_iu_i}\bar{u}_j)-\frac{\partial}{\partial x_j}(\overline{pu_j}-\bar{p}\bar{u}_j) \\
		&+\frac{\partial}{\partial x_j}(\nu \frac{\partial k_{sgs}}{\partial x_j})+\frac{\partial}{\partial x_j}(\tau_{ij}\bar{u}_i)\\
		&-\nu(\overline{\frac{\partial u_i}{\partial x_j}\frac{\partial u_i}{\partial x_j}}-\frac{\partial \bar{u}_i}{\partial x_j}\frac{\partial \bar{u}_i}{\partial x_j})-\tau_{ij}\frac{\partial \bar{u}_i}{\partial x_j}.
	\end{split}
\end{equation}

To further simplify this equation, we use the analogy proposed by \citet{knight1998}:
\begin{equation}
	\frac{1}{2}\left(\overline{u_iu_iu_j}-\overline{u_iu_i}\bar{u}_j\right) = \frac{1}{2} \left(\overline{u_i u_i u_k}-\bar{u}_i \bar{u}_i \bar{u}_k-2 k_{sgs} \bar{u}_k\right) \approx \tau_{i k} \bar{u}_i,
\end{equation}
and we noticed that with the continuity equation, the term representing diffusion by pressure effect can be taken as
\begin{equation}
	\frac{\partial}{\partial x_j}(\overline{pu_j}-\bar{p}\bar{u}_j) = \overline{u_j\frac{\partial p}{\partial x_j}}-\bar{u}_j\frac{\partial \bar{p}}{\partial x_j}.
\end{equation}
Finally, the simplified SGS kinetic energy equation can be written as:
\begin{equation}
	\begin{split}
		\frac{\partial k_{sgs}}{\partial t} = &-\frac{\partial}{\partial x_j}(k_{sgs}\bar{u}_j)-\left(\overline{u_j\frac{\partial p}{\partial x_j}}-\bar{u}_j\frac{\partial \bar{p}}{\partial x_j}\right) \\
		&+\frac{\partial}{\partial x_j}(\nu \frac{\partial k_{sgs}}{\partial x_j})-\nu(\overline{\frac{\partial u_i}{\partial x_j}\frac{\partial u_i}{\partial x_j}}-\frac{\partial \bar{u}_i}{\partial x_j}\frac{\partial \bar{u}_i}{\partial x_j})-\tau_{ij}\frac{\partial \bar{u}_i}{\partial x_j}.
	\end{split}
\end{equation}

In the simulation of incompressible channel flow, which is our test case, the non-dimensional equation can be obtained by using half channel width $h$ and wall shear velocity $u_{\tau}$:
\begin{equation}
	\begin{split}
		\frac{\partial k_{sgs}}{\partial t} = &-\frac{\partial}{\partial x_j}(k_{sgs}\bar{u}_j)-\left(\overline{u_j\frac{\partial p}{\partial x_j}}-\bar{u}_j\frac{\partial \bar{p}}{\partial x_j}\right) \\
		&+\frac{1}{Re_{\tau}} \frac{\partial^2 k_{sgs}}{\partial x_j \partial x_j}-\frac{1}{Re_{\tau}}(\overline{\frac{\partial u_i}{\partial x_j}\frac{\partial u_i}{\partial x_j}}-\frac{\partial \bar{u}_i}{\partial x_j}\frac{\partial \bar{u}_i}{\partial x_j})-\tau_{ij}\frac{\partial \bar{u}_i}{\partial x_j},\label{eq:ksgs}
	\end{split}
\end{equation}
where $Re_{\tau} = \frac{\rho u_{\tau} h}{\nu}$ is the wall shear Reynolds number and $\bar{u} = \bar{u}^*/u_{\tau}$ is actually the dimensionless velocity $u^+$.

In equation~(\ref{eq:ksgs}), there are three quantities that need to be modelled, they are the SGS stress tensor $\tau_{ij}$, the pressure dilatation $\Pi_p$ and the viscous dissipation $\epsilon$, respectively:
\refstepcounter{equation}
$$
	\tau_{ij}, \quad \Pi_p =\left(\overline{u_j\frac{\partial p}{\partial x_j}}-\bar{u}_j\frac{\partial \bar{p}}{\partial x_j}\right), \quad \varepsilon = (\overline{\frac{\partial u_i}{\partial x_j}\frac{\partial u_i}{\partial x_j}}-\frac{\partial \bar{u}_i}{\partial x_j}\frac{\partial \bar{u}_i}{\partial x_j}).
	\eqno{(\theequation{\mathit{a},\mathit{b},\mathit{c}})}
$$
Precise and accurate modelling of these unclosed quantities is crucial to the success of the new model. In the upcoming section, we will present modelling approaches for these unclosed quantities.
\subsection{The quasi-dynamic process and joint-constraint model}
In this section, we will start by introducing the quasi-dynamic process and helicity flux dual channel constraint for incompressible flows and then extend the result to compressible flows.

A series expansion of filtered quantity is given by \citet{bedford1993conjunctive} as:
\begin{equation}
	\begin{aligned}
		\overline{f g}-\bar{f} \bar{g}= & 2 \alpha \frac{\partial \bar{f}}{\partial x_k} \frac{\partial \bar{g}}{\partial x_k}+\frac{1}{2 !}(2 \alpha)^2 \frac{\partial^2 \bar{f}}{\partial x_k \partial x_l} \frac{\partial^2 \bar{g}}{\partial x_k \partial x_l} \\
		                                & +\frac{1}{3 !}(2 \alpha)^2 \frac{\partial^3 \bar{f}}{\partial x_k \partial x_l \partial x_m} \frac{\partial^2 \bar{g}}{\partial x_k \partial x_l \partial x_m}+\cdots,
	\end{aligned}
\end{equation}
where
\begin{equation}
	\alpha(y)=\int_{-\infty}^{\infty} x^2 G(x, y) \mathrm{d} x.
\end{equation}
For a box filter, an exact value $\alpha = \Delta^2/24$ can be obtained. In practice, this is set to be $2\alpha = C_0\Delta_k^2$, where $\Delta_k$ is the grid length scale in $k$ direction. Thus, the unclosed terms can be modelled, viz.
\begin{equation}
	\tau_{ij} = C_0\Delta_k^2\frac{\partial \bar{u}_i}{\partial x_k}\frac{\partial \bar{u}_j}{\partial x_k}+\frac{1}{2!}(C_0^2\Delta_k^2\Delta_l^2)\frac{\partial^2 \bar{u}_i}{\partial x_k\partial x_l}\frac{\partial^2 \bar{u}_j}{\partial x_k \partial x_l} + \cdots,
\end{equation}

As it has been proved by previous successful applications of QKM model \citep{qi2022a}, it is enough for us to reserve only the first term for modelling $\tau_{ij}:$
\begin{equation}
	\tau_{ij}\approx C_0\Delta_k^2\frac{\partial \bar{u}_i}{\partial x_k}\frac{\partial \bar{u}_j}{\partial x_k}.\label{eq:tau}
\end{equation}
As the SGS kinetic energy $k_{sgs} = \frac{1}{2}\tau_{kk}$, one can obtain
\begin{equation}
	C_0 = \frac{2k_{sgs}}{\Delta_l^2\dfrac{\partial \bar{u}_k}{\partial x_l}\dfrac{\partial \bar{u}_k}{\partial x_l}}.
\end{equation} 
For the unclosed quantities $\Pi_p$ and $\varepsilon$, we also apply this method to get
\begin{eqnarray}
	\Pi_p &\approx& C_0\Delta_l^2\frac{\partial \bar{p}}{\partial x_l}\frac{\partial \bar{u}_k}{\partial x_l \partial x_k}, \\
	\varepsilon & \approx& C_0\Delta_k^2\frac{\partial^2 \bar{u}_i}{\partial x_j\partial x_k}\frac{\partial^2 \bar{u}_i}{\partial x_j\partial x_k}.
\end{eqnarray}

Now eqaution~(\ref{eq:ksgs}) is closed and the value of $C_0$ can be solved dynamically. Through this so-called quasi-dynamic process, a more precise distribution of $k_{sgs}$ can be obtained. As mentioned, the deconvolution form of equation~(\ref{eq:tau}) shows a high correlation with the real value but is often numerically unstable. 
To maintain a strong correlation with the real values while ensuring numerical stability, we use the equation for SGS kinetic energy to derive precise SGS kinetic energy and determine the appropriate coefficients for the deconvolution model. We employ a joint-constraint approach to restrict the Smagorinsky model, which exhibits excellent numerical stability.
% To achieve balance between accuracy and stability, we use the $k_{sgs}$ obtained from the quasi-dynamic process and the helicity dual channel flux to constrain the Smagorinksy model to get better behaviour on depict physical process such as intermittency and energy cascade.

To tighten the constraints of the Smagorinsky model, the methodology leverages the kinetic energy flux and helicity flux. The primary objective is to ensure that the modified model demonstrates a strong correlation with both fluxes while maintaining numerical stability. Thus, the proposed approach employs a least square constraint method to determine the appropriate coefficient for the Smagorinsky model.

We denote the quantity related to the modified Smagorinksy model with a dynamic constant $C_{sm}$ with superscript $SM$ and quantities related to the deconvolution model with a dynamic coefficient $C_0$ that is solved through SGS kinetic energy equation with superscript $EQ$. The kinetic energy flux and dual-channel helicity flux can be expressed as
\begin{eqnarray}
	\Pi_{E}^{SM} =C_{sm}\Delta^2|\bar{S}|\bar{S}_{ij}\frac{\partial \bar{u}_i}{\partial x_j},\quad \Pi_{E}^{EQ} &=& C_0\Delta_k^2\frac{\partial \bar{u}_i}{\partial x_k}\frac{\partial \bar{u}_j}{\partial x_k}\frac{\partial \bar{u}_i}{\partial x_j},\\
	\Pi_{H1}^{SM} = -C_{sm}\Delta^2|\bar{S}|\bar{S}_{ij}\frac{\partial \bar{\omega}_i}{\partial x_j}, \quad \Pi_{H1}^{EQ} &=&  -C_0\Delta_k^2\frac{\partial \bar{u}_i}{\partial x_k}\frac{\partial \bar{u}_j}{\partial x_k}\frac{\partial \bar{\omega}_i}{\partial x_j},\\
	\Pi_{H2}^{SM} =  -\nabla\times (C_{sm}\Delta^2|\bar{S}|\bar{S}_{ij})\frac{\partial \bar{u}_i}{\partial x_j},\quad \Pi_{H2}^{EQ} &=& -C_0\Delta_k^2\left(\frac{\partial \bar{u}_j}{\partial x_k}\frac{\partial \bar{\omega}_i}{\partial x_k}-\frac{\partial \bar{u}_i}{\partial x_k}\frac{\partial \bar{\omega}_j}{\partial x_k}\right)\frac{\partial \bar{u}_i}{\partial x_j}.
\end{eqnarray}
To get the constraint for $C_{sm}$, the variation of the model coefficient is frozen. We can make the assumption that $\nabla C_{sm}$ is negligible so that:
\begin{equation}
	\nabla\times (C_{sm}\Delta^2|\bar{S}|\bar{S}_{ij})\frac{\partial \bar{u}_i}{\partial x_j} = C_{sm} \nabla\times (\Delta^2|\bar{S}|\bar{S}_{ij})\frac{\partial \bar{u}_i}{\partial x_j}.
\end{equation}
Now the deviation between the Smagorinsky model and the dynamic deconvolution model can be expressed as
\begin{equation}
	\begin{aligned}
		\delta_E(C_{sm}) &= \left(\Pi_{E}^{EQ} - \Pi_{E}^{SM}\right)^2,\\
		\delta_{H1}(C_{sm}) &= \left(\Pi_{H1}^{EQ} - \Pi_{H1}^{SM}\right)^2,\\
		\delta_{H2}(C_{sm}) &= \left(\Pi_{H2}^{EQ} - \Pi_{H2}^{SM}\right)^2.\\
	\end{aligned}
\end{equation}
Since the total derivation $\delta = \delta_{E} + \delta_{H1} + \delta_{H2}$ can achieve minimum in a reasonable giving range of $C_{sm}$, take $\partial \delta/{\partial C_{sm}} = 0$, we got the optimized $C_{sm}$:
\begin{eqnarray}
	C_{sm} &=& \frac{\Pi_{E}^{EQ}B_1 + \Pi_{H1}^{EQ}B_2 + \Pi_{H2}^{EQ}B_3}{B_1^2+B_2^2+B_3^2},\label{eq:sm-incomp}\\
	B_1 &=& -\Delta^2|\bar{S}|\bar{S}_{ij}\frac{\partial \bar{u}_i}{\partial x_j}, \label{eq:constrain}\\
	B_2 &=& -\Delta^2|\bar{S}|\bar{S}_{ij}\frac{\partial \bar{\omega}_i}{\partial x_j}, \\
	B_3 &=& -\nabla \times (\Delta^2|\bar{S}|\bar{S}_{ij})\frac{\partial \bar{u}_i}{\partial x_j}.
\end{eqnarray}
With equation~(\ref{eq:sm-incomp}), the eddy viscosity can be obtained, i.e. 
\begin{equation}
	\nu_{sgs} = C_{sm}\Delta^2|\tilde{S}|.
\end{equation}
For the range of clipping for $C_{sm}$, the relation recommend by \citet{garnier2009large} is adopted, i.e. 
\begin{eqnarray}
	&\nu_{sgs} + \nu \geqslant 0,\\
	&C_{sm} \leqslant C_{max},
\end{eqnarray}
where $C_{max}$ can be taken as the square of Smagorinksy constant $0.18^2$.

\subsection {Application in incompressible rotating channel flow}
\begin{figure}
	\centering
	\includegraphics[width=0.7\linewidth]{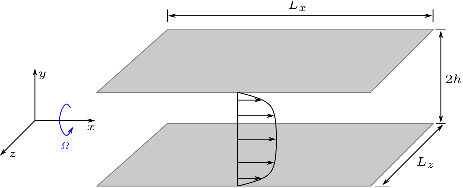}
	\caption{Schematic diagram of streamwise-rotating channel}
	\label{fig:channel}
\end{figure}

As the first test case of the new model, incompressible streamwise-rotating channel flow is widely studied in literature\citep{yang2018,oberlack2006group,alkishriwi2008large}.  Figure \ref{fig:channel} shows the schematic diagram for the streamwise-rotating channel flow. 

% The continuity and momentum equations for the DNS of this case can be written as

% \begin{subeqnarray}
% 	\nabla \cdot \vect{u} &=& 0, \slabel{eq:continuity-channel} \\[3pt]
% 	\frac{\partial \vect{u}}{\partial t} + \vect{u}\cdot \nabla \vect{u} + 2 \vect{\Omega} \times \vect{u} &=& -\nabla p + \nu \nabla^2\vect{u} - \frac{\vect{\Pi}}{\rho}. \slabel{eq:momentum-channel}
% \end{subeqnarray}

The rotating vector of the system is defined as $\vect{\Omega} = \left(\Omega_x, 0, 0\right)$, where $\Omega_x$ is set to be a constant so that the rotation is homogeneous. The flow is driven by constant streamwise pressure gradient $\vect{\Pi} = \left(\Pi_x, 0, 0\right)$ and sufficient computation time (approximately $100h/u_{\tau}$) is taken to ensure the time duration is long enough for achieving a statistically stationary state. Periodic boundary conditions are applied to the streamwise and spanwise of the computational domain and the no-slip and impermeable boundary conditions are applied to the top and bottom walls. For LES, the dynamic Smagorinsky model (DSM) and the wall-adapting local eddy-viscosity (WALE) model are compared with the new model. The test filter width of the DSM model is chosen to be $2\Delta$, where $\Delta = (\Delta_x\Delta_y\Delta_z)^{1/3}$ is the length scale of the mesh. For the WALE model, the model constant is chosen to be $C_w = \sqrt{10.6}C_s$ following \citet{ducros1998wall}, where $C_s$ is the Smagorinksy model constant. 

As it is vitally important to control the numerical dissipation in LES\citep{chapelier2017}, a high-resolution pseudo-spectral method is used to solve the control equation. The grid near the wall is refined by using Cheyshev-Gauss-Lobatto points and flow variables are expanded into Chebyshev polynomials in the wall norm direction. Fourier series are used to expand flow variables in the streamwise and spanwise direction on uniform grids. More computational details can be found in the literature. To examine the effectiveness of the new model, \textit{a priori} tests were carried out for channel flow with Reynolds number ($Re_{\tau} = u_{\tau}h/\nu$) fixed at 180 and the rotation number ($Ro_{\tau}=2\Omega h/u_{\tau}$) fixed at 7.5. Then \textit{a posteriori} analysis of the performance of the new model in higher Reynolds number ($Re_{\tau} = 395$) is also presented.
Table \ref{tab:grid} shows the grid settings for direct numerical simulation (DNS) and LES in two different Reynolds numbers. The computational domain is set to $32\pi \times 2 \times 8\pi$ in the streamwise, wall-normal, and spanwise direction, this computational domain is large enough to capture large-scale intermittency and most energetic eddy motions including Taylor-G{\"o}tler vortices. 

\begin{table}
	\begin{center}
  \def~{\hphantom{0}}
	\begin{tabular}{lccccc}
		  Case& $N_x\times N_y \times N_z$   &  $\Delta x^{+}$ & $\Delta y_{min}^{+}$ & $\Delta y_{max}^{+}$ & $\Delta z^{+}$ \\[3pt]
		 DNS-Re180   & $1024\times 128 \times 512$ & 17.67 & 0.11& 4.42 & 8.84\\
		 DNS-Re395   & $4096 \times 192 \times 1536$ & 9.69 & 0.11 & 6.46& 6.46\\
		 LES-Re180   & $256\times 96 \times 128$ & 70.69 & 0.19 & 5.89& 35.34\\
		 LES-Re395   & $512\times 96 \times 256$ & 77.56 & 0.42 & 12.93& 38.78\\
	\end{tabular}
	\caption{Grid settings and grid resolutions of the simulations in the incompressible streamwise rotating channel flow at $Re=180$ and $Re=395$}
	\label{tab:grid}
	\end{center}
  \end{table}

\subsubsection{\textit{A priori} tests}
For the \textit{a priori} tests, filtered quantities are obtained from DNS data by applying a Gaussian filter with a filter width equal to $4\Delta$. The filtering process was conducted in spectral space and the transfer function takes the form 
\begin{equation}
	\hat{G}(\kappa) = \exp\left(\frac{-\bar{\Delta}^2\kappa^2}{24}\right),
\end{equation}
where $\bar{\Delta}$ is the filter width in one direction. Note that filtering is only adopted in the streamwise and spanwise directions.
The correlation coefficient is defined as
\begin{equation}
	\beta=\frac{cov(R,M)}{\sigma(R)\sigma(M)},
\end{equation}
where $R$ is the real quantity from DNS data and $M$ is the modelled quantity. $cov(X,Y)$ denotes the covariance between quantities $X$ and $Y$ and $\sigma(X)$ is the standard deviation of the distribution of the quantity $X$. Here, we examine correlations of SGS stress tensor $\tau_{11}, \tau_{12}$, SGS kinetic energy $k_{sgs} = \frac{1}{2}\tau_{kk}$ and SGS dissipation $\Pi_{\Delta}^{E} = \tau_{ij}\frac{\partial \bar{u}_i}{\partial x_j}$.

\begin{figure}
	\centering
	\includegraphics[width=\linewidth]{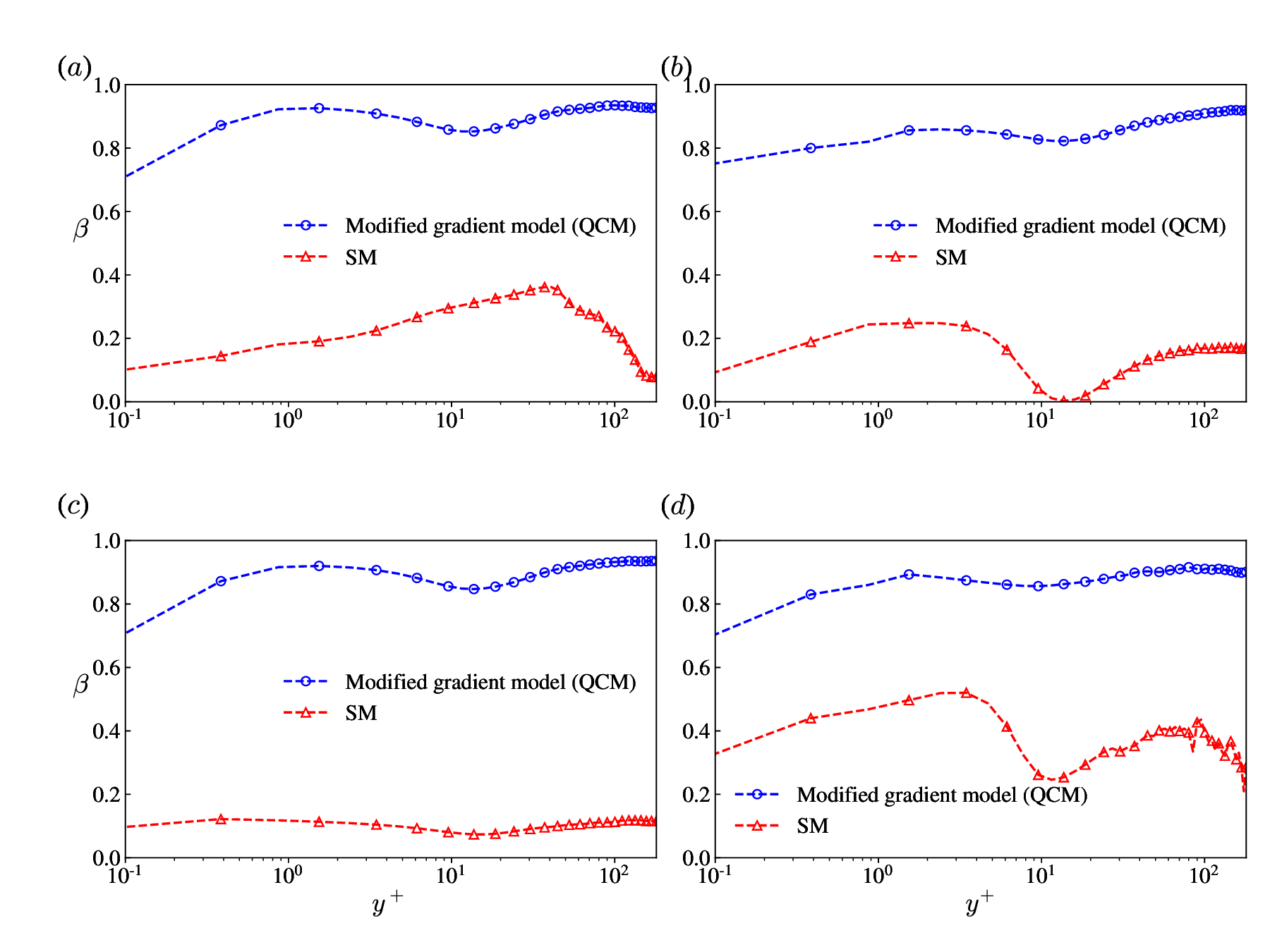}
	\caption{Correlations of modified gradient model using QCM dynamic coefficient and the Smagorinksy model with respect to filtered DNS of incompressible rotating channel flow at $Re_{\tau} = 180, Ro_{\tau}=7.5$: (a) the component of SGS stress $\tau_{11}$; (b) the component of SGS stress $\tau_{12}$; (c) the SGS kinetic energy $k_{sgs}$; (d) the subgrid dissipation $\Pi_{\Delta}^{E}$.}
	\label{fig:re180-priori}
\end{figure}

Figure \ref{fig:re180-priori} shows that all the modelled quantities from QCM have rather high correlations with real values than the SM model, almost all correlation coefficients maintain above 0.8 along the wall-normal direction. More importantly, the correlation coefficient of SGS kinetic energy is significantly improved compared to the SM model, which has a poor correlation with the filtered DNS data. We can estimate that the new model has better performance in depicting subgrid-scale kinetic motion and obtaining the correct magnitude of SGS kinetic energy.

\begin{figure}
	\centering
	\includegraphics[width=0.7\linewidth]{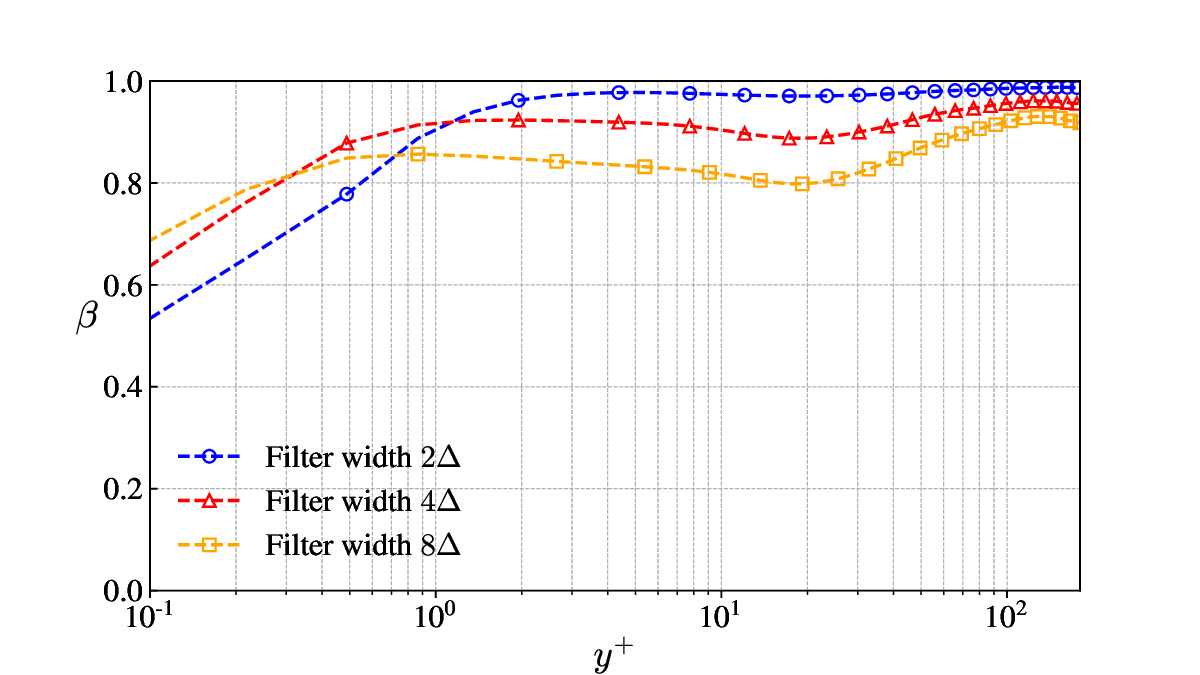}
	\caption{Correlation coefficients of helicity flux at three different filter widths.}
	\label{fig:re180-priori-heli}
\end{figure}

Helicity flux $\Pi_{\Delta}^{H} = -\tau_{ij}\bar{R}_{ij} - \gamma_{ij}\bar{\Omega}_{ij}$ is a new quantity introduced in this study as a constraint of LES model. To further examine the correlation coefficients at different filter widths, \textit{a priori} tests of helicity flux were carried out with filter width equal to $2\Delta, 4\Delta, 8\Delta$, separately. Figure \ref{fig:re180-priori-heli} shows that even at a large filter width, the helicity flux from the QCM model has high similarity with the real data.

At the \textit{a priori} test level, QCM shows good performance, but it should be noted that these correlations are not necessarily a measure of the alignment between simulated and exact quantities such as subgrid-scale stresses. Next, we will discuss the results of the \textit{a posteriori} tests in incompressible streamwise rotating channel flow at $Re=180$ and $Re=395$ with the rotation number $Ro=7.5$.

\subsubsection{\textit{A posteriori} tests}
\begin{figure}
	\centering
	\includegraphics[width=\linewidth]{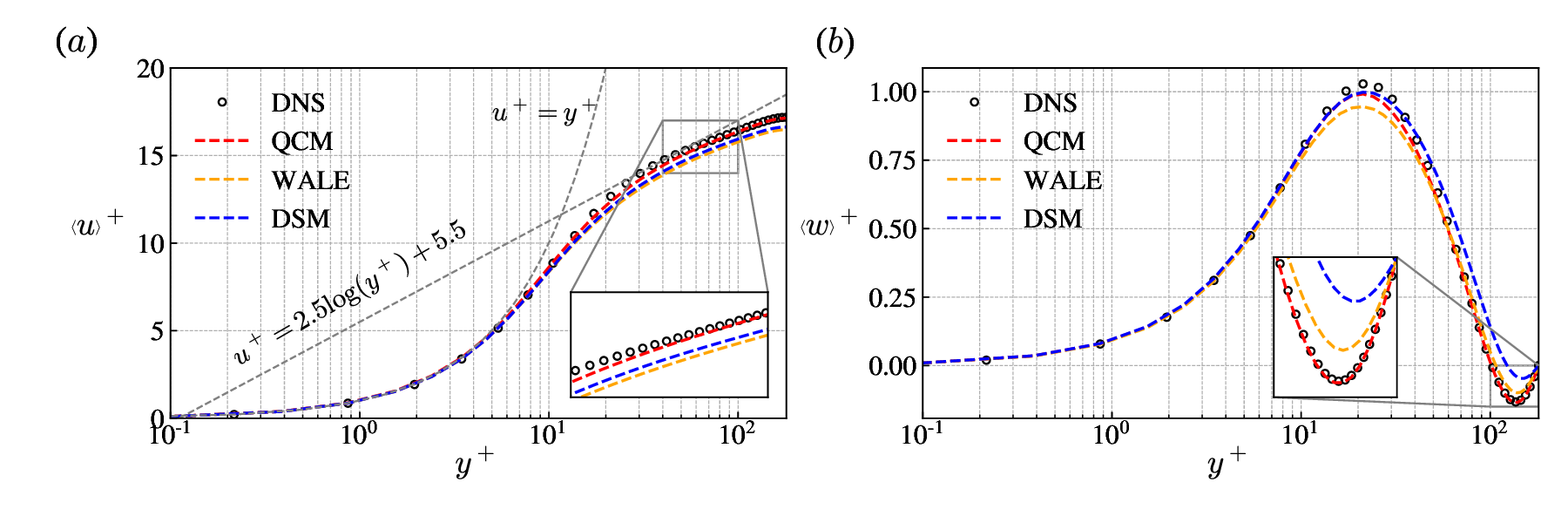}
	\caption{Profiles of mean velocity, note that every two values have been marked for visual clarity for DNS data: (a) streamwise mean velocity; (b) spanwise mean velocity.}
	\label{fig:re180-yplus}
\end{figure}

The profiles of the streamwise and spanwise mean velocity obtained from DNS, the QCM, the DSM, and the WALE model are compared in figure \ref{fig:re180-yplus}. Averaging is performed over the $x-z$ plane, which is marked by $\langle\cdot \rangle$. As it is discussed by previous study \citep{yu2022}, the effect of streamwise rotation is reflected as a smaller streamwise velocity, making the streamwise velocity profile deviate from the log law. Moreover, on the spanwise velocity profile, there exists reverse flow in the core regions and distinct secondary flows near the wall. It can be observed from figure \ref{fig:re180-yplus} that results from the QCM are better than those of the other two models. The DSM and the WALE model underpredicted the mean streamwise velocity in the centre region and the DSM failed to capture reverse flows accurately, the WALE model did not give the correct peak of secondary flow. In contrast, the QCM is capable of capturing both the reverse and the secondary flows more accurately.

\begin{figure}
	\centering
	\includegraphics[width=\linewidth]{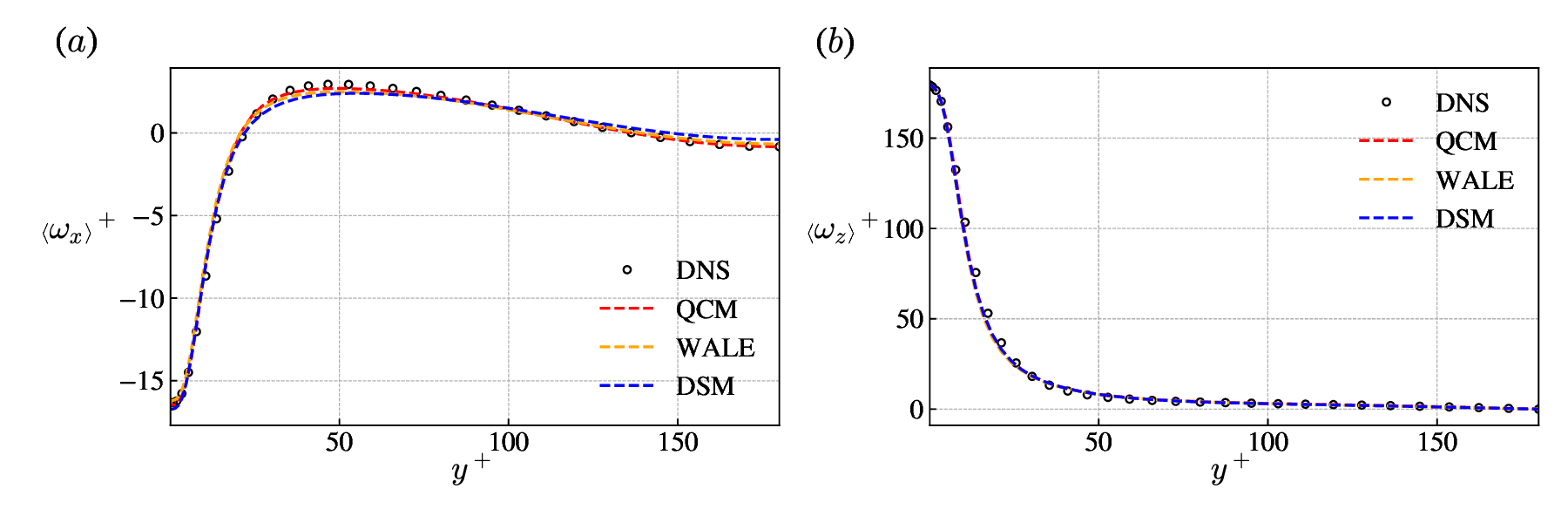}
	\caption{The profiles of mean vorticity in the streamwise and spanwise directions.}
	\label{fig:re180-vort}
\end{figure}

According to the homogeneity conditions in the streamwise and spanwise directions, non-zero mean vorticity in the streamwise direction is produced due to the existence of mean secondary flow, making the streamwise vorticity a main characteristic of streamwise rotating turbulent channel flows. In figure \ref{fig:re180-vort}, we can see results of the three LES models are similar, except the QCM model is slightly better at predicting the streamwise vorticity near the wall.

\begin{figure}
	\centering
	\includegraphics[width=0.7\linewidth]{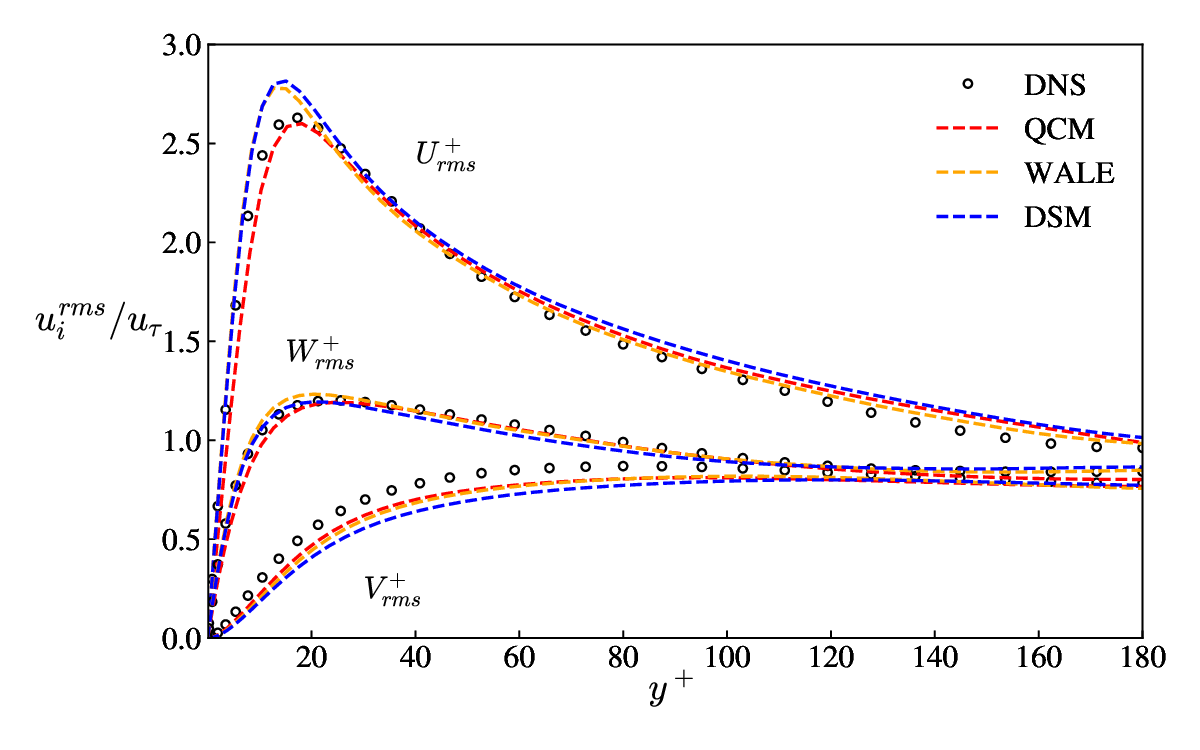}
	\caption{Wall-normal profiles of turbulent intensities from QCM, DSM, and WALE model in channel flow with streamwise rotation. Dots represent DNS data, every two value has been marked for visual clarity for DNS data.}
	\label{fig:re180-reynolds}
\end{figure}

For LES simulations, the mean streamwise component of intensity in the near-wall region is expected to be smaller with respect to unfiltered DNS data in order to be consistent with the filtered DNS. In figure \ref{fig:re180-reynolds}, it can be observed that the QCM gives a better prediction of turbulent intensities while the streamwise turbulent intensity by the other two models is significantly higher than the unfiltered DNS data. This is a qualitative indicator that the new model is improved at the near-wall region.

\begin{figure}
	\centering
	\includegraphics[width=\linewidth]{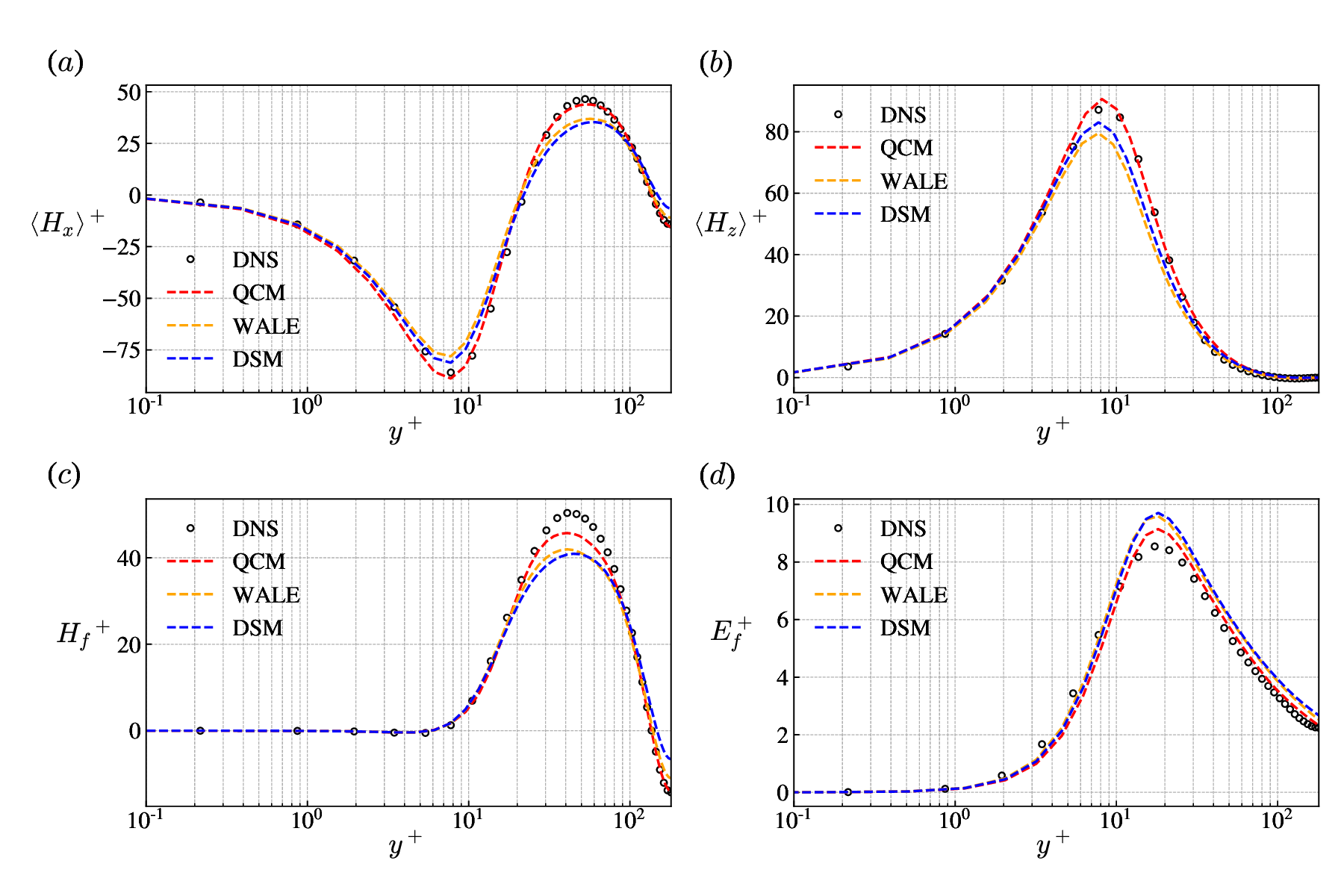}
	\caption{Wall normal profiles of helicity and resolved kinetic energy: (a) streamwise component of mean helicity; (b) spanwise component of mean helicity; (c) fluctuating helicity; (d) fluctuating kinetic energy}
	\label{fig:re180-heli}
\end{figure}

In wall-bounded turbulent flows, new energy transfer paths exist compared to homogeneous and isotropic flows, hence it is necessary to investigate the capability of predicting it for LES models. We can define the mean and fluctuating energy as 
\begin{equation}
	E m=\langle u\rangle^2+\langle w\rangle^2, \quad E f=u^{\prime 2}+v^{\prime 2}+w^{\prime 2}.
\end{equation}
Similarly, the mean and fluctuating helicity can be defined as
\begin{equation}
	H m=H x m+H z m=\langle u\rangle\left\langle\omega_x\right\rangle+\langle w\rangle\left\langle\omega_z\right\rangle, \quad H f=u^{\prime} \omega_x^{\prime}+v^{\prime} \omega_y^{\prime}+w^{\prime} \omega_z^{\prime} .
\end{equation}
Note that there exist another two components of mean and fluctuating fields according to the Schwarz inequality. Their ensemble averages are zero, so they are absent in the above decomposition. Figure \ref{fig:re180-heli} shows the profile of mean helicity in the streamwise and spanwise direction, together with the profile of fluctuating helicity and fluctuating energy. From figure \ref{fig:re180-heli}, we can see that the QCM yields a precise prediction compared with the DNS, but the other models still have an obvious deviation from the real value, especially around the peak.

\begin{figure}
	\centering
	\includegraphics[width=\linewidth]{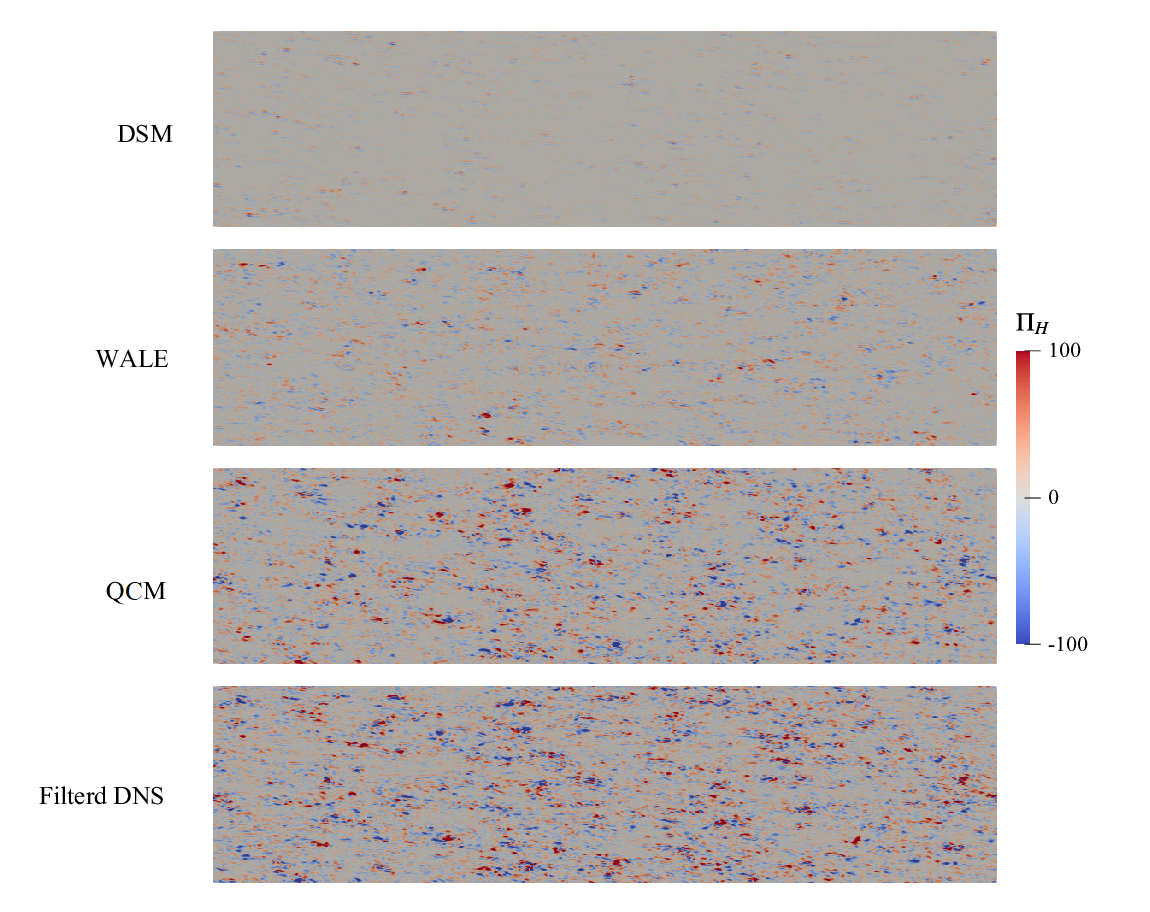}
	\caption{Helicity flux distribution on the wall, the filtered DNS data is obtained by applying Gaussian filter with filter width $4\Delta$}
	\label{fig:re180-pih}
\end{figure}

Figure \ref{fig:re180-pih} shows the helicity flux distribution on the wall, it is clear that the QCM gives similar distribution compared to the filtered DNS data, while the result of the other two models lacks details. The existence of more intense helicity flux fluctuation also suggests stronger energy backscatter in the near-wall region.

\begin{figure}
	\centering
	\includegraphics[width=\linewidth]{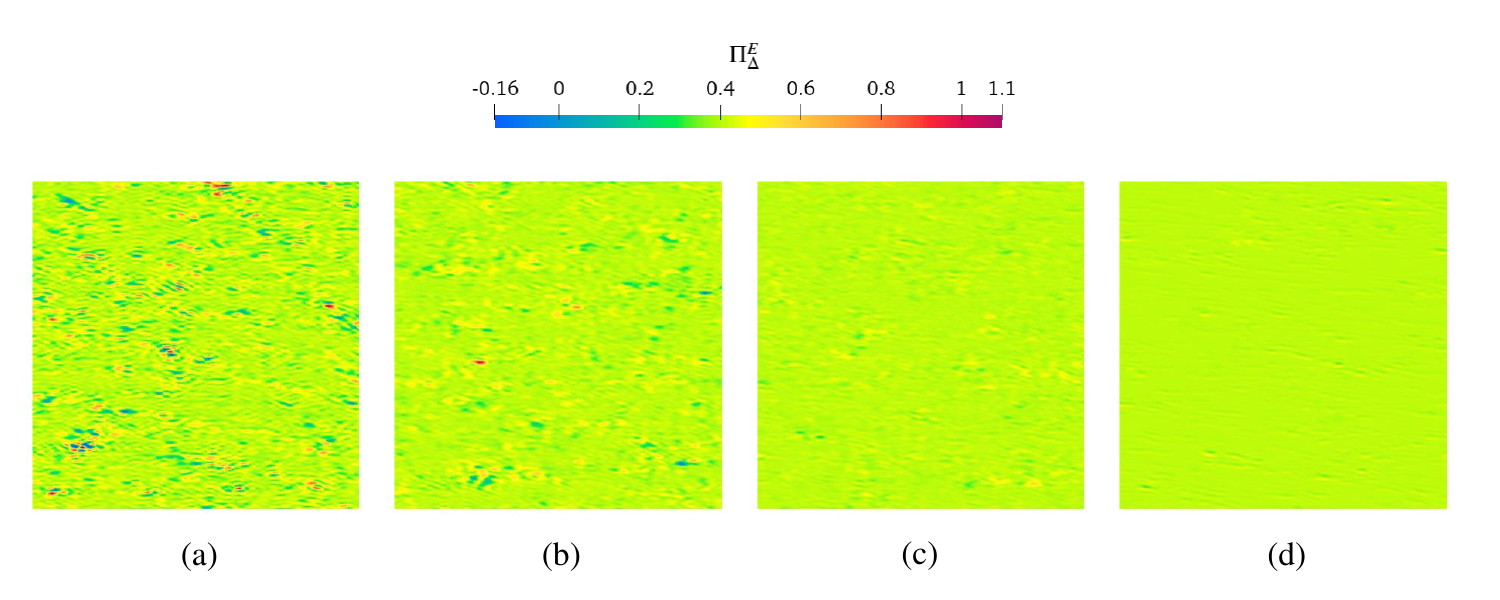}
	\caption{kinetic energy flux distribution on the wall, the filtered DNS data is obtained by applying Gaussian filter with filter width $4\Delta$. (a) Filtered DNS; (b) QCM; (c) DSM; (d) WALE.}
	\label{fig:re180-piE}
\end{figure}

As can be seen in Figure \ref{fig:re180-piE}, the QCM gives similar SGS KEF distribution compared to the filtered DNS data, there exist several spots of negative KEF. The WALE model can not predict negative KEF hence it is deviated from the real distribution, the DSM can supply negative KEF, but its distribution shows a low correlation with the real data.

\begin{figure}
	\centering
	\includegraphics[width=\linewidth]{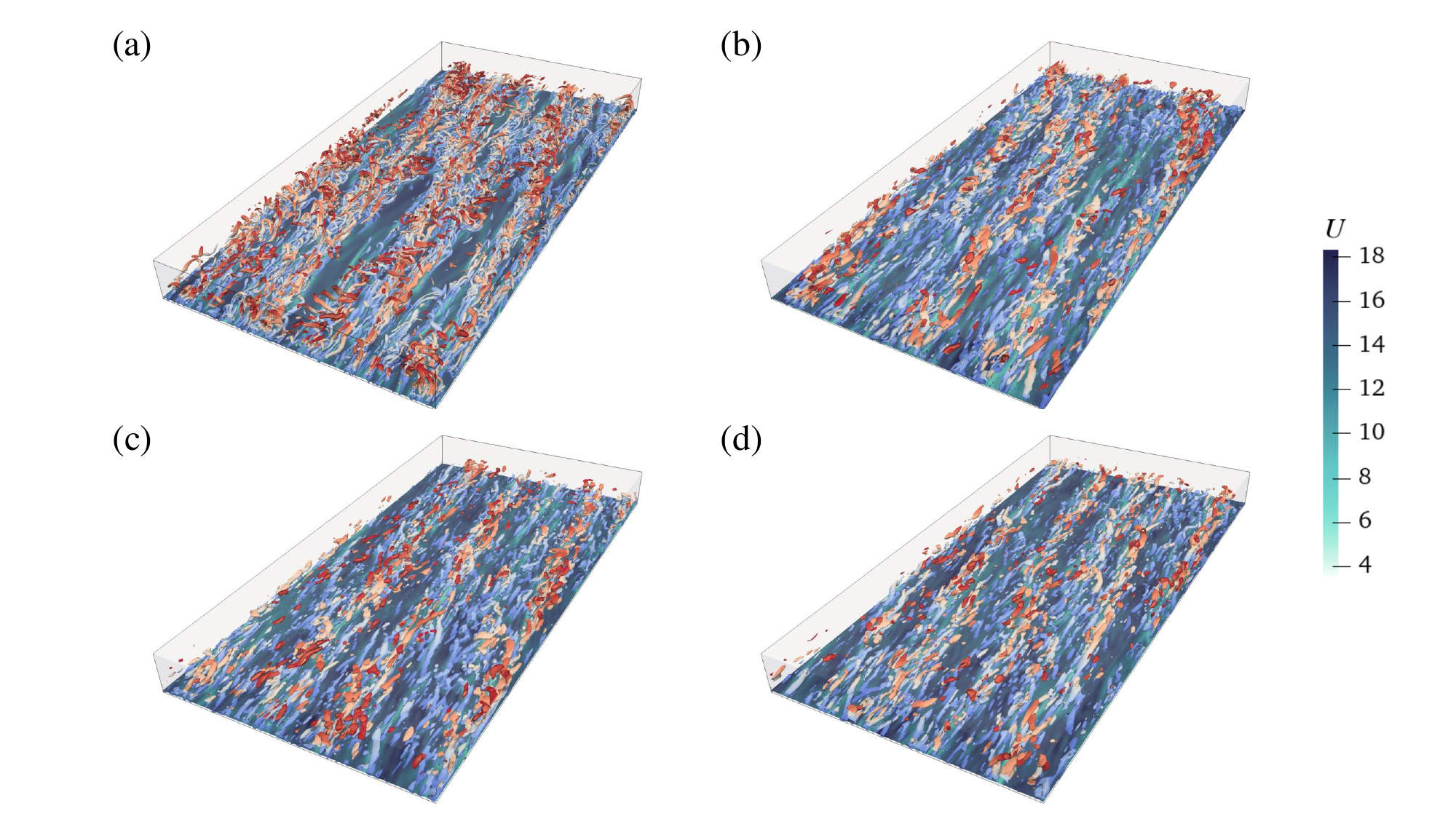}
	\caption{Instantaneous isosurface of Q (second invariant of the strain-rate tensor, Q=70) obtained from (a) DNS, (b) the QCM, (c) the WALE model, (d) the DSM of incompressible turbulent channel flow, only the lower half is shown.}
	\label{fig:re180-Q}
\end{figure}

The coherent structure is one of the key components of turbulent flows. Figure \ref{fig:re180-Q} shows the instantaneous isosurface of Q, which is the second invariant of the strain-rate tensor. Here, $Q=70$ is chosen as a threshold value. From figure \ref{fig:re180-Q}, we can see that in the near-wall region, all three models show a large-scale strip structure, while the QCM can predict more abundant coherent structures, especially in the near-wall region. We have to note that according to the Q isosurface, even if the mesh is much coarser than the DNS mesh and so much information of small-scale vortices is lost, the QCM model can still give rather accurate prediction of turbulence quantity distributions. This is encouraging that it shows the new model is improved in representing the effect of small to medium-scale vortices.

\begin{figure}
	\centering
	\includegraphics[width=\linewidth]{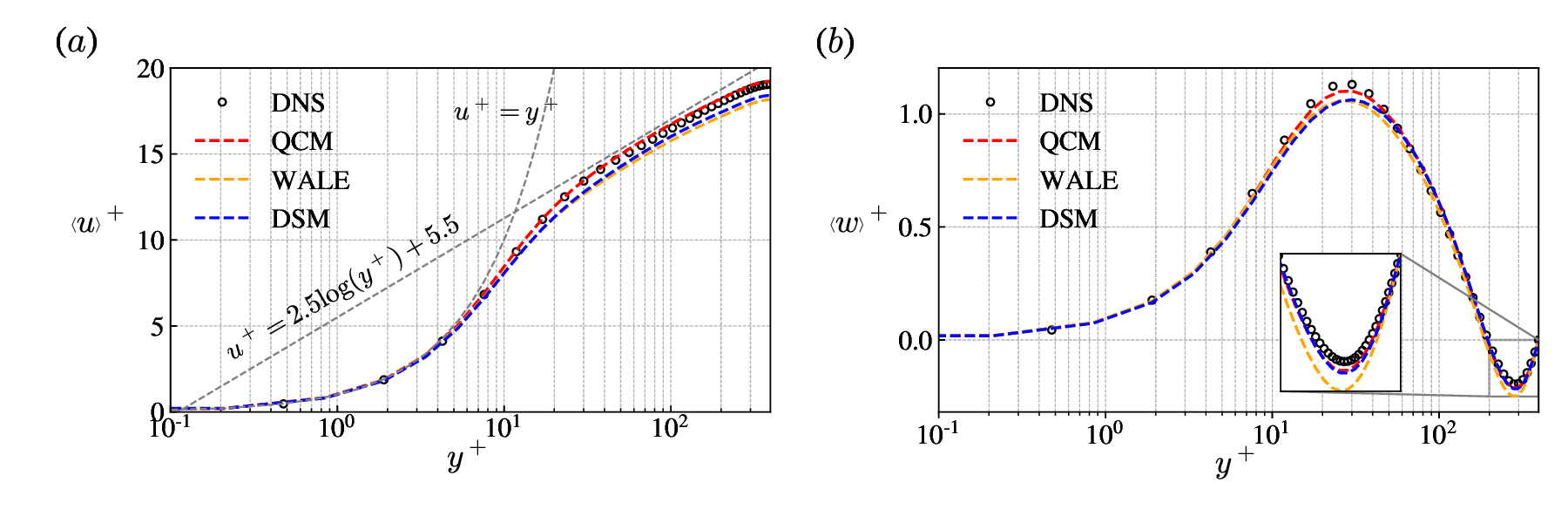}
	\caption{Profiles of mean helicity in $Re_{\tau}=395$ channel flow, note that every three values have been marked for visual clarity for DNS data: (a) streamwise mean velocity; (b) spanwise mean velocity.}
	\label{fig:re395-yplus}
\end{figure}

\begin{figure}
	\centering
	\includegraphics[width=\linewidth]{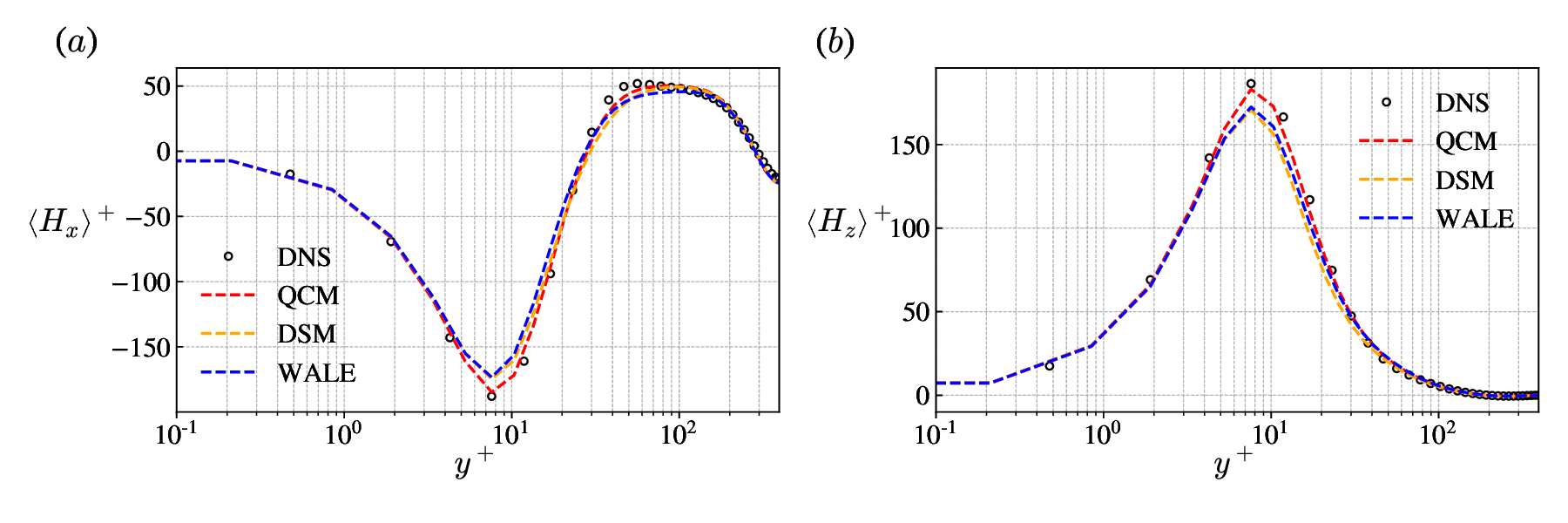}
	\caption{Profiles of mean velocity in $Re_{\tau}=395$ channel flow, note that every three values have been marked for visual clarity for DNS data: (a) streamwise mean helicity; (b) spanwise mean helicity.}
	\label{fig:re395-heli}
\end{figure}

To further validate the new model, another test case of streamwise rotating channel flow with Reynolds number $Re_{\tau}=395$, rotation number $Ro=7.5$ was conducted. The computational domain is the same as the $Re_{\tau}=180$ case. Figure \ref{fig:re395-yplus} shows the streamwise and spanwise mean velocity profiles, normalized by $u_{\tau}$. Figure \ref{fig:re395-heli} shows the mean helicity distribution in both streamwise and spanwise directions. We can see that the QCM still shows better performance compared with the other two models. The helicity distribution from the QCM is highly consistent with the DNS data. 

\section{LES modelling for compressible flows}
\subsection{Governing equations for compressible LES modelling}
For the compressible flow simulation, applying a filter on the Navier-Stokes equation gives
\begin{eqnarray}
	\frac{\partial \bar{\rho}}{\partial t} + \frac{\partial \bar{\rho}\tilde{u}_i}{\partial x_j} &=& 0,\\
	\frac{\partial \bar{\rho} \tilde{u}_i}{\partial t}+\frac{\partial \bar{\rho} \tilde{u}_i \tilde{u}_j}{\partial x_j}&=&-\frac{\partial \bar{p}}{\partial x_i}+\frac{\partial \tilde{\sigma}_{i j}}{\partial x_j}-\frac{\partial \tau_{i j}}{\partial x_j}, \\
	\frac{\partial \bar{\rho} \tilde{E}}{\partial t}+\frac{\partial(\bar{\rho} \tilde{E}+\bar{p}) \tilde{u}_j}{\partial x_j}&=&-\frac{\partial \tilde{q}_j}{\partial x_j}+\frac{\partial \tilde{\sigma}_{i j} \tilde{u}_i}{\partial x_j}-\frac{\partial C_p Q_j}{\partial x_j}-\frac{\partial J_j}{\partial x_j},
\end{eqnarray}

where $\left(\tilde{\cdot}\right)$ represents density-weighted (Favre) filtering ($\tilde{\phi} = \bar{\rho}\phi/\bar{\rho}$). Compare to the filtered equation of incompressible flow, there are several differences: The energy equation is needed here and the filtered total energy is written as 

\refstepcounter{equation}
$$
\bar{\rho}\tilde{E} = \bar{\rho}C_v\tilde{T}+\frac{1}{2}\bar{\rho}\tilde{u}_i\tilde{u}_i + \bar{\rho}k_{sgs}, \quad
\bar{\rho}k_{sgs}=\frac{1}{2}\bar{\rho}(\widetilde{u_iu_i}-\tilde{u}_i\tilde{u}_i)
\eqno{(\theequation{\mathit{a},\mathit{b}})}.
$$
The resolved heat flux $\tilde{q}_j$, the SGS heat flux $Q_j$ and the SGS turbulent diffusion term $J_j$ are expressed as
\begin{eqnarray}
\tilde{q}_j &=& \frac{C_p\mu(\tilde{T})}{Pr}\frac{\partial \tilde{T}}{\partial x_j},\\
Q_j &=& \bar{\rho}(\widetilde{u_jT}-\tilde{u}_j\tilde{T}),\\
J_j &=& \frac{1}{2}\bar{\rho}(\widetilde{u_iu_iu_j}-\widetilde{u_iu_i}\tilde{u}_j),
\end{eqnarray}
where the molecular viscosity $\mu$ takes from Sutherland's law.

\subsection{Dual channels of helicity flux and the SGS kinetic energy equation for compressible flows}
The dual channels of helicity in compressible flows can be derived in a similar way as in incompressible flows. We can obtain the governing equation of large-scale helicity $\tilde{H} = \tilde{\vect{u}}\cdot \tilde{\vect{\omega}}$ as follows:
\begin{equation}
	\frac{\partial \tilde{H}}{\partial t}+\frac{\partial \tilde{J}_j}{\partial x_j}=-\Pi_{\Delta}^{H1}-\Pi_{\Delta}^{H2}+\Phi_l^1+\Phi_l^2+D_l^1+D_l^2,
\end{equation}
where $\tilde{J}_j$ is the spatial transport term and $\Phi_l$ is the pressure term, $D_l$ is the viscosity term. Here, superscript 1 is named the first channel originating from the Farve filtered velocity governing equation, and superscript 2 is named the second channel originating from the Farve filtered vorticity equation. Here, we mainly focus on the dual channels of helicity flux. The first and second channels of helicity flux are written respectively as
\begin{eqnarray}
	\Pi_{\Delta}^{H1} &=& -\bar{\rho} \tilde{\tau}_{i j} \frac{\partial}{\partial x_j}\left(\frac{\tilde{\omega}_i}{\bar{\rho}}\right), \\
	\Pi_{\Delta}^{H2} &=& -\epsilon_{i k l} \frac{\partial \bar{\rho} \tilde{\tau}_{l j}}{\partial x_k} \frac{\partial}{\partial x_j}\left(\frac{\tilde{u}_i}{\bar{\rho}}\right)-\bar{\rho} \tilde{\tau}_{i j} \frac{\partial}{\partial x_j}\left(\epsilon_{i k l} \tilde{u}_k \frac{\partial}{\partial x_l}\left(\frac{1}{\bar{\rho}}\right)\right) .
\end{eqnarray}
If we take the filtered density as constant, the equations above would recover to the form of incompressible flows. For the kinetic energy flux, it has the form same as in incompressible 
flows, i.e. 
\begin{equation}
\Pi_{\Delta}^E = \tau_{ij}\tilde{S}_{ij}.
\end{equation}

Similar to the SGS kinetic energy equation of incompressible flow, one can derive the SGS kinetic equation of compressible flow, written as
\begin{equation}
	\begin{aligned}
		\frac{\partial \bar{\rho} k_{s g s}}{\partial t} = &-\frac{\partial \bar{\rho} k_{s g s} \tilde{u}_j}{\partial x_j}  - \tau_{ij}\tilde{S}_{ij} -\varepsilon_s - \varepsilon_d + \Pi_p\\
		&-\frac{\partial J_j}{\partial x_j} + \frac{\partial}{\partial x_j}\left( \bar{\mu}\frac{\partial k_{sgs}}{\partial x_j}\right) + \frac{\partial }{\partial x_j}\left[ \tau_{ij}\tilde{u}_i + \bar{\mu}\frac{\partial}{\partial x_i}\left(\frac{\tau_{ij}}{\bar{\rho}}\right) +R\tilde{q}_j\right], \label{eq:ksgs-compressible}
	\end{aligned}
\end{equation}
with the following notation:
\begin{eqnarray}
		\varepsilon_s&=&2 \bar{\mu}\left[\widetilde{\mathbb{S}_{i j} \mathbb{D}_{i j}}-\widetilde{\mathbb{S}}_{i j} \widetilde{\mathbb{D}}_{i j}\right],\\
		\varepsilon_d&=&\frac{\partial}{\partial x_j}\left[\frac{5}{3}\left(\bar{\mu} \widetilde{u_j \frac{\partial u_k}{\partial x_k}}-\bar{\mu} \tilde{u}_j \frac{\partial \tilde{u}_k}{\partial \tilde{x}_k}\right)\right], \\
		\Pi_p&=&\overline{p \frac{\partial u_k}{\partial x_k}}-\bar{p} \frac{\partial \tilde{u}_k}{\partial x_k} .
\end{eqnarray}
Where $R = C_p-C_v$ is the specific gas constant and $\tilde{\mathbb{S}}_{ij}, \tilde{\mathbb{D}}_{ij}$ are the traceless strain rate/velocity gradient tensor. Again, like the incompressible SGS kinetic energy equation, the equation of compressible SGS kinetic energy is not fully closed, several terms need to be modelled: the SGS stress tensor $\tau_{ij}$, the solenoidal dissipation $\varepsilon_s$, the dilatational dissipation $\varepsilon_d$ and the pressure dilatation $\Pi_p$. These terms will be modelled using the quasi-dynamic process method introduced in the next section.

\subsection{The joint-constraint model for compressible flows}
As is mentioned in the derivation of the quasi-dynamic method of incompressible flow, by taking series expansion of the unclosed quantities, we can get the approximation of them, expressed as
\begin{eqnarray}
	\tau_{ij} &\approx& C_0\Delta_k^2\bar{\rho}\frac{\partial \tilde{u}_i}{\partial x_k}\frac{\partial \tilde{u}_j}{\partial x_k}, \label{eq:tau-comp}\\
	\varepsilon_s &\approx& 2C_0\Delta_k^2\mu(\tilde{T})\frac{\partial \tilde{\mathbb{S}}_{ij}}{\partial x_k}\frac{\partial \tilde{\mathbb{D}}_{ij}}{\partial x_k},  \\ 
	\varepsilon_d &\approx& \frac{5}{3}\frac{\partial}{\partial x_j}\left[C_0\Delta_l^2\mu(\tilde{T})\frac{\partial \tilde{u}_j}{\partial x_l}\frac{\partial^2 \tilde{u}_k}{\partial x_k \partial x_l}\right], \\
	\Pi_p &\approx& C_0\Delta_l^2\frac{\partial \bar{p}}{\partial x_l}\frac{\partial^2 \tilde{u}_k}{\partial x_l \partial x_k}.
\end{eqnarray}
Here we note that all four unclosed quantities use the same $C_0$, this has been proved reasonable by both \textit{a prior} and \textit{a posteriori} tests \citep{qi2022a}.

From equation~(\ref{eq:tau-comp}), we can find that
\begin{equation}
	\tau_{kk} = C_0\Delta_l^2\bar{\rho}\frac{\partial \tilde{u}_k}{\partial x_l}\frac{\partial \tilde{u}_k}{\partial x_l},
\end{equation}
and we have the relation $\tau_{kk} = 2\bar{\rho}k_{sgs}$, so we can confirm the real value of $C_0$ as 
\begin{equation}
	C_0 = \frac{2k_{sgs}}{\Delta^2_l\frac{\partial \tilde{u}_k}{\partial x_l}\frac{\partial \tilde{u}_k}{\partial x_l}}.
\end{equation}

The kinetic energy flux $\Pi_{\Delta}^{E} = \tau_{ij}\tilde{S}_{ij}$, is another quantity used to constrain the model coefficient. Using eqaution~(\ref{eq:tau-comp}), we can derive the deconvolution model representation of these quantities as follows:
\begin{eqnarray}
	\Pi_{\Delta}^{E} &=& C_0\Delta_k^2\bar{\rho}\frac{\partial \tilde{u}_i}{\partial x_k}\frac{\partial \tilde{u}_j}{\partial x_k}\tilde{S}_{ij}, \\
	\Pi_{\Delta}^{H1} &=& -\bar{\rho} C_0\Delta_k^2\bar{\rho}\frac{\partial \tilde{u}_i}{\partial x_k}\frac{\partial \tilde{u}_j}{\partial x_k} \frac{\partial}{\partial x_j}\left(\frac{\tilde{\omega}_i}{\bar{\rho}}\right),
\end{eqnarray}
\begin{equation}
	\begin{aligned}
		\Pi_{\Delta}^{H2} = &-\epsilon_{i k l} \frac{\partial \bar{\rho}}{\partial x_k}\left( C_0\Delta_m^2\bar{\rho}\frac{\partial \tilde{u}_l}{\partial x_m}\frac{\partial \tilde{u}_j}{\partial x_m}\right) \frac{\partial}{\partial x_j}\left(\frac{\tilde{u}_i}{\bar{\rho}}\right) \\
		& -\bar{\rho} C_0\Delta_k^2\bar{\rho}\frac{\partial \tilde{u}_i}{\partial x_k}\frac{\partial \tilde{u}_j}{\partial x_k} \frac{\partial}{\partial x_j}\left(\epsilon_{i k l} \tilde{u}_k \frac{\partial}{\partial x_l}\left(\frac{1}{\bar{\rho}}\right)\right).
	\end{aligned}
\end{equation}
Again, adopting the least-square method, the constraint for the Smagorinksy model coefficient can be obtained:
\begin{eqnarray}
	C_{sm} &=& \frac{\Pi_{E}^{EQ}B_1 + \Pi_{H1}^{EQ}B_2 + \Pi_{H2}^{EQ}B_3}{B_1^2+B_2^2+B_3^2}, \label{eq:sm}\\
	B_1 &=& -2\bar{\rho}\Delta^2|\tilde{S}|\tilde{\mathbb{S}}_{ij}\tilde{S}_{ij}, \\
	B_2 &=& -\bar{\rho}^2\Delta^2|\tilde{S}|\tilde{\mathbb{S}}_{ij}\frac{\partial \tilde{\omega}_i/\bar{\rho}}{\partial x_j}, \\
	B_3 &=& -\nabla \times (\bar{\rho}\Delta^2|\tilde{S}|\tilde{\mathbb{S}}_{ij})\frac{\partial \tilde{u}_i/\bar{\rho}}{\partial x_j} + \bar{\rho}^2\Delta^2|\tilde{S}|\tilde{\mathbb{S}}_{ij}\frac{\partial}{\partial x_j}\left(\epsilon_{ikl}\tilde{u}_k\frac{\partial}{\partial x_l}\left( \frac{1}{\bar{\rho}}\right)\right).
\end{eqnarray}
With the new coefficient provided by equation~(\ref{eq:sm}), the anisotropic part of the Smagorinksy model can be determined, i.e. 
\begin{equation}
	\tau_{ij}^{mod} - \frac{1}{3}\delta_{ij}\tau_{kk}^{mod} = -2\bar{\rho}C_{sm}\Delta^2 |\tilde{S}|\tilde{\mathbb{S}_{ij}}.
\end{equation}
The isotropic part of the Smagorinksy model can be obtained directly from the relation $\tau_{kk}^{mod}=2\bar{\rho}k_{sgs}$. 

Another problem to deal with in LES modelling of compressible flow is the model for SGS heat flux, here we can apply the series expansion method to get
\begin{equation}
	Q_j = C_0\Delta_k^2\bar{\rho}\frac{\partial \tilde{u}_j}{\partial x_k}\frac{\partial \tilde{T}}{\partial x_k}.
\end{equation}
The SGS Prandtl number $Pr_{sgs}$, can be obtained through the constraint $\partial Q^{mod}_j/\partial x_j = \partial Q_j/\partial x_j$, where $Q^{mod}_j = -\frac{\mu_{sgs}}{Pr_{sgs}}\frac{\partial \tilde{T}}{\partial x_j}$ is the commonly used diffusion model \citep{Moin1991}. Therefore, we have 
\begin{equation}
	Pr_{sgs}=-\frac{\partial\left(v_{sgs} \frac{\partial \tilde{T}}{\partial x_j}\right) / \partial x_j}{\partial\left(C_0 \Delta_k^2 \frac{\partial \tilde{u}_j}{\partial x_k} \frac{\partial \tilde{T}}{\partial x_k}\right) / \partial x_j},
\end{equation}
where $\nu_{sgs} = \mu_{sgs}/\bar{\rho}$.

So far, we have constructed the constraint model QCM based on the kinetic energy flux and dual-channel helicity flux for compressible flows. We will examine the effectiveness of the new model in the next section. The test case is chosen to be the transonic flow in a rotating annular pipe and the hypersonic flow over a rotating cone.

\subsection {Application in compressible rotating annular pipe flow}
\begin{figure}
	\centering
	\includegraphics[width=0.7\linewidth]{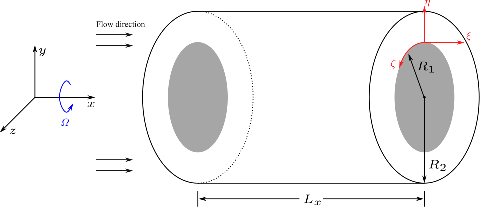}
	\caption{Schematic diagram of streamwise-rotating annular pipe}
	\label{fig:pipe}
\end{figure}

The first compressible test case for the new model is Mach 0.8 flow over an annular pipe with streamwise rotation. Figure \ref{fig:pipe} shows the schematic diagram of this case. The size of the computational domain is $L_x\times R_1 \times R_2 = 20\times 1 \times 3$, the Mach number is $Ma = 0.8$, the bulk Reynolds number is 6550 and the friction Reynolds number $\overline{Re}_{\tau} = \bar{u}_{\tau}\delta/\nu$ is 513.5 ($\delta = (R_2-R_1)/2$ is the half-width of the annular pipe and $\nu$ is the kinetic viscosity, $Re_{\tau}=513.5$ is the DNS result). We have to note that due to the rotation effect, the friction velocity $u_{\tau}$ is different for the inner wall and the outer wall, we use the averaged value of wall friction velocity $\bar{u}_{\tau}$ to calculate the friction Reynolds number. The Prandtl number $Pr=\mu C_p\kappa$ is 0.7, where $\kappa$ is the thermal conductivity and $C_p$ is the specific heat at constant pressure. The flow is driven at a constant mass flow rate to match the mass flow from the DNS. Periodic boundary conditions are applied in the streamwise direction and the no-slip boundary condition and fixed temperature boundary condition are applied at the wall, i.e. $T_{w} = 288.15 \rm{K}$. The solver used is a fidelity finite difference solver. For DNS, a mixed scheme that combines the seventh-order upwind scheme and the seventh-order WENO scheme is used for spatial discretization of the convective term. A shock wave sensor is used as an indicator to switch from a linear scheme to a WENO-type scheme in the non-smooth region to ensure numerical stability and high precision. For LES, a six-order central difference scheme is adopted for spatial discretization of the convective term to minimize numerical dissipation. For both DNS and LES, a six-order central difference scheme is adopted for the viscous term and the third-order Runge-Kutta scheme is used for time integration. The centrifugal force and the Coriolis force are utilized in the system by adding explicit source terms in the Navier-Stokes equations, i.e. 
\begin{subeqnarray}
	\frac{\partial \vect{U}}{\partial t} &+& \nabla \cdot \vect{F}_c(\vect{U}) = \nabla \cdot \vect{F}_v(\vect{U}) + \vect{S}_{\omega}, \slabel{eq:compressible} \\[3pt]
	\vect{U} &=& \left[ \rho, \rho u, \rho v, \rho w, \rho E\right]^T, \\[3pt]
	\vect{S}_{\omega} &=& \left[0,0,\rho(\omega^2y+2\omega w), \rho(\omega^2z-2\omega v), \rho \omega^2(vy+wz)\right]^T.
\end{subeqnarray}

Where $\vect{U}$ is the vector of conservative variables and $S_{\omega}$ is the source term that drives system rotation. $\vect{F}_c(\vect{U})$ and $\vect{F}_v(\vect{U})$ are the convective term and viscous term, respectively. For this case, we take $\omega = 0.4 \rm{rad/s}$ and the rotation number $Ro = \frac{2\Omega R_1}{\bar{u}_{\tau}}$ is about 15. The dynamic Smagorinksy model and the WALE model are chosen to compare with the new model, the grid settings and grid resolutions of the simulations are listed in Table \ref{tab:grid-pipe}.

\begin{table}
	\begin{center}
  \def~{\hphantom{0}}
	\begin{tabular}{lcccccc}
		  Case& $N_{\xi}\times N_{\eta}\times N_{\zeta}$   &  $\Delta \xi^{+}$ & $\Delta \eta_{min}^{+}$ & $\Delta \eta_{max}^{+}$ & $\Delta \zeta_{min}^{+}$ &$\Delta \zeta_{max}^{+}$ \\[3pt]
		 DNS   & $512\times 256 \times 512$ & 9.84 & 0.29& 4.03 & 3.09 &9.28\\
		 LES   & $256 \times 128 \times 256$ & 19.5 & 0.57 & 7.74& 6.13 &18.38\\
	\end{tabular}
	\caption{Grid settings and grid resolutions of the simulations in the compressible streamwise rotating annular pipe flow. Note that the grid resolution is based on the average of inner and outer wall friction velocity.}
	\label{tab:grid-pipe}
	\end{center}
\end{table}

\begin{figure}
	\centering
	\includegraphics[width=0.7\linewidth]{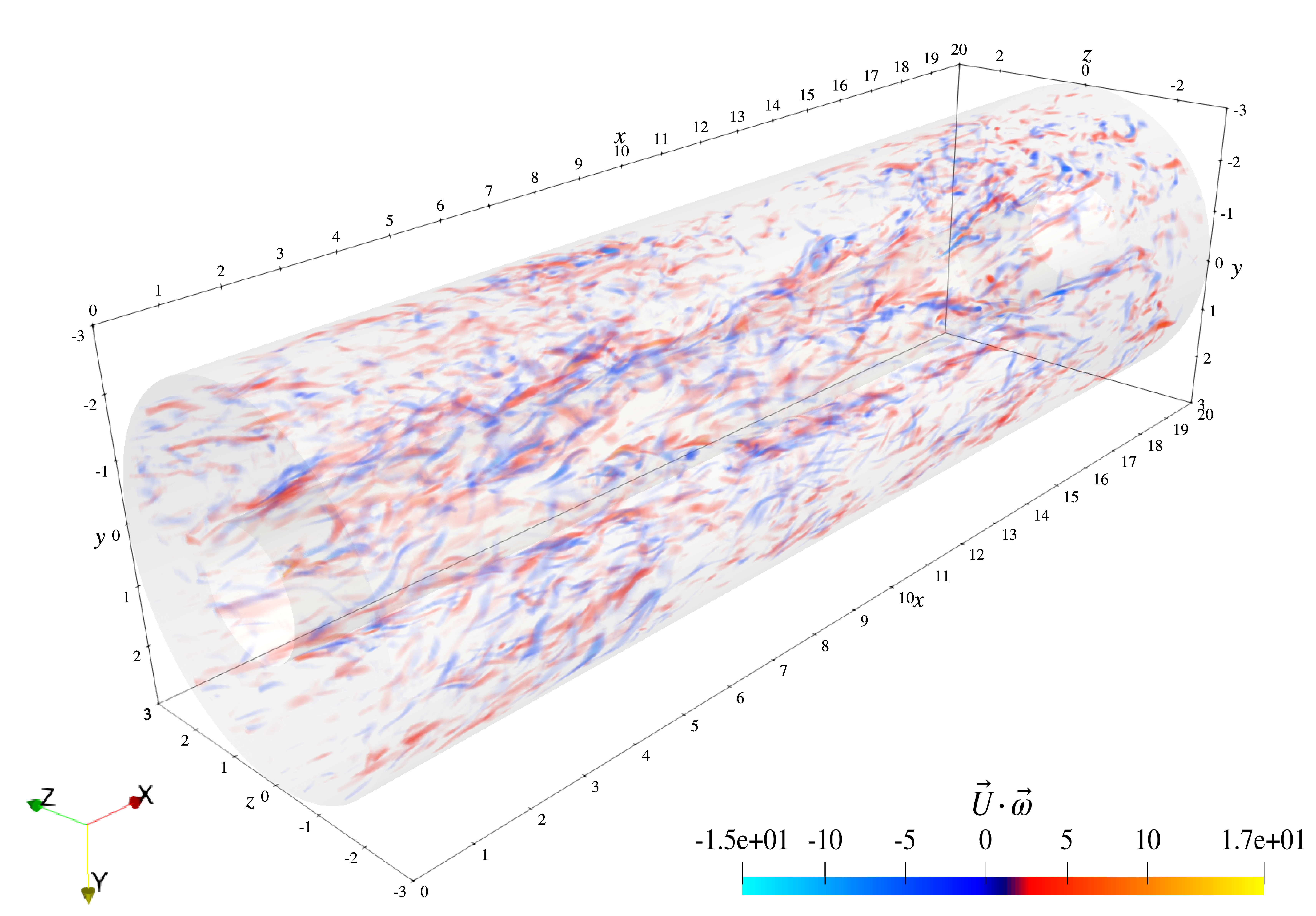}
	\caption{Helicity distribution of streamwise-rotating annular pipe}
	\label{fig:pipe-heli}
\end{figure}

Figure \ref{fig:pipe-heli} depicts the helicity distribution of the streamwise-rotating annular pipe, as we can see, abundant helical structures exist in this figure. Compared to the incompressible rotating channel case, this problem is more challenging due to a more sophisticated energy transfer process between fluid and the system, hence an elaborated modelling process is needed.

\begin{figure}
	\centering
	\includegraphics[width=\linewidth]{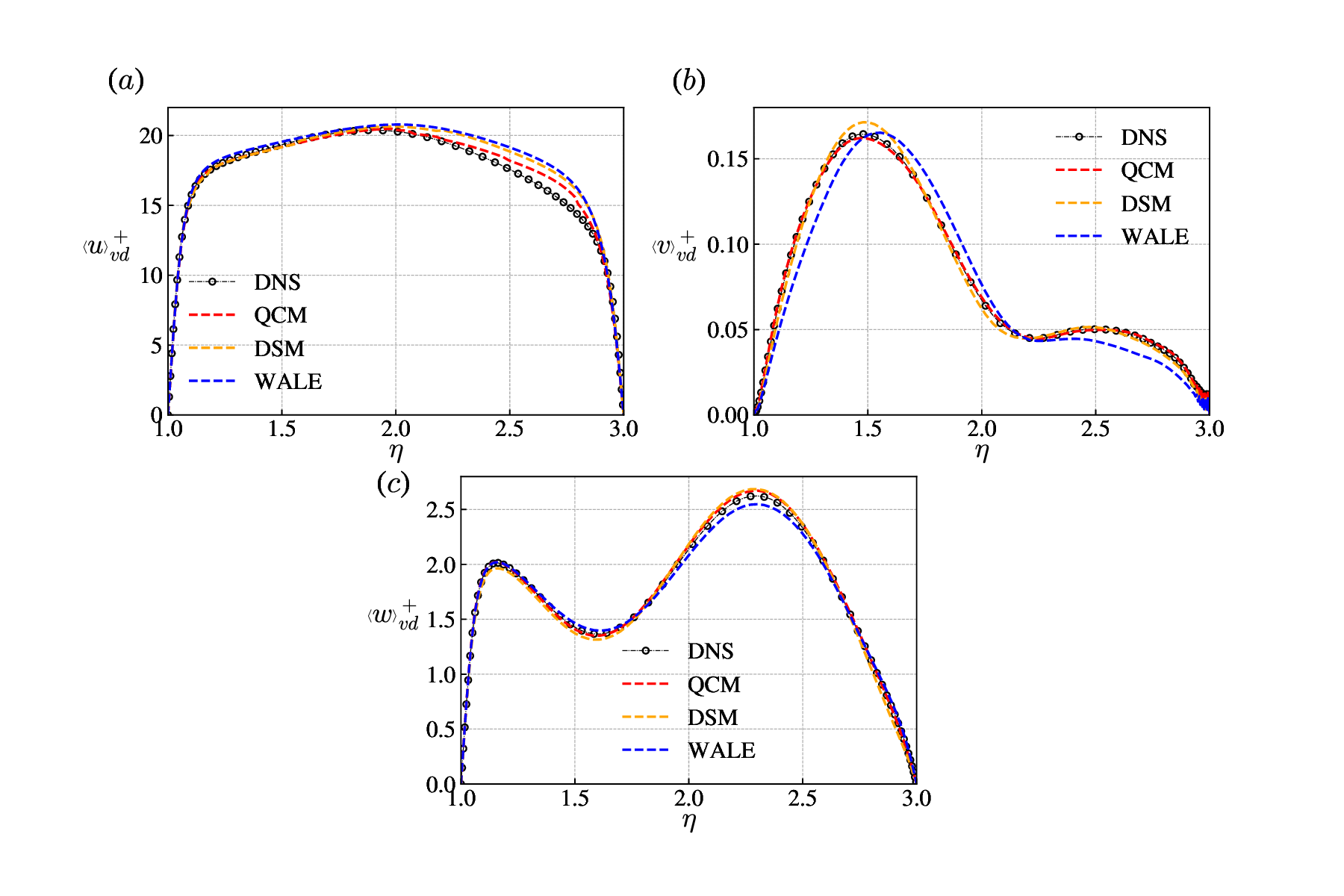}
	\caption{Mean velocity profiles of the compressible rotating annular pipe flow}
	\label{fig:pipe-U}
\end{figure}

First, we compare profiles of mean velocities. Figure \ref{fig:pipe-U} shows the profile of streamwise mean velocity, circumferential mean velocity, and mean velocity in the wall-normal direction. The van-Driest transformation is adopted here:
\begin{equation}
	\langle u\rangle_{vd}^{+} = \int_{0}^{u^+}\sqrt{\langle \rho/\rho_w \rangle}d\langle u\rangle^{+}.
\end{equation}
In the following study, we will use $\langle u\rangle_{vd}^{+}, \langle v\rangle_{vd}^{+}, \langle w\rangle_{vd}^{+}$ to represent the van Driest transformed velocity in streamwise, circular and wall-normal directions respectively.
As we can see, although all three models tend to over-predict the mean streamwise velocity near the outer wall, the QCM's performance is better than DSM and WALE models. For the wall-normal velocity, the WALE model completely deviated from the DNS data and DSM tends to over-predict the peak value near the inner wall. For the circumferential velocity, the DSM and the QCM model give similar predictions and the WALE model under-predicted the peak value. Generally speaking, the QCM gives a more accurate prediction of mean velocity profiles in this case.

\begin{figure}
	\centering
	\includegraphics[width=0.8\linewidth]{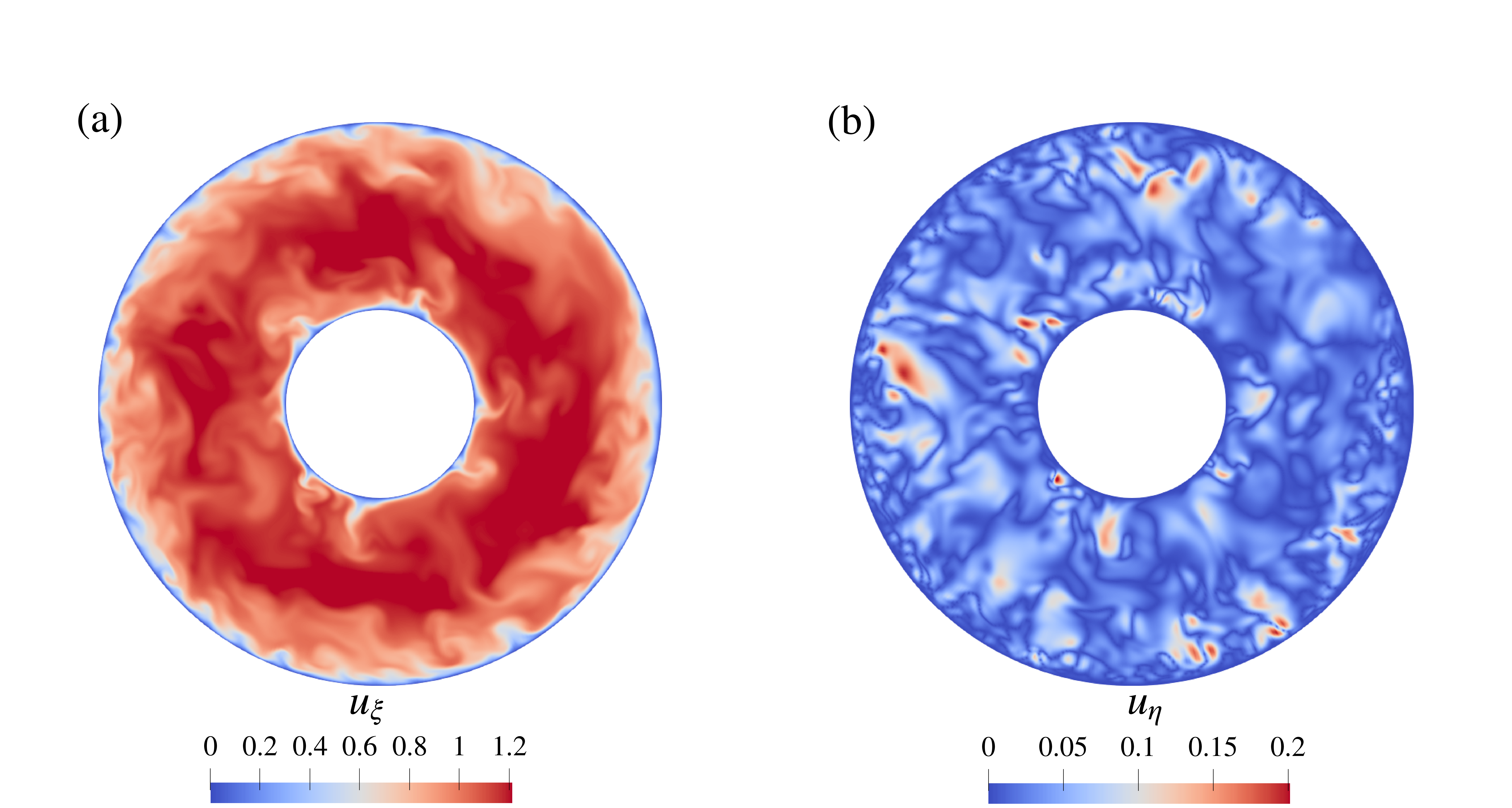}
	\caption{Velocity distribution of DNS data on a streamwise tangential plane. (a) streamwise velocity; (b) wall-normal velocity}
	\label{fig:pipe-U-dns}
\end{figure}

\begin{figure}
	\centering
	\includegraphics[width=0.9\linewidth]{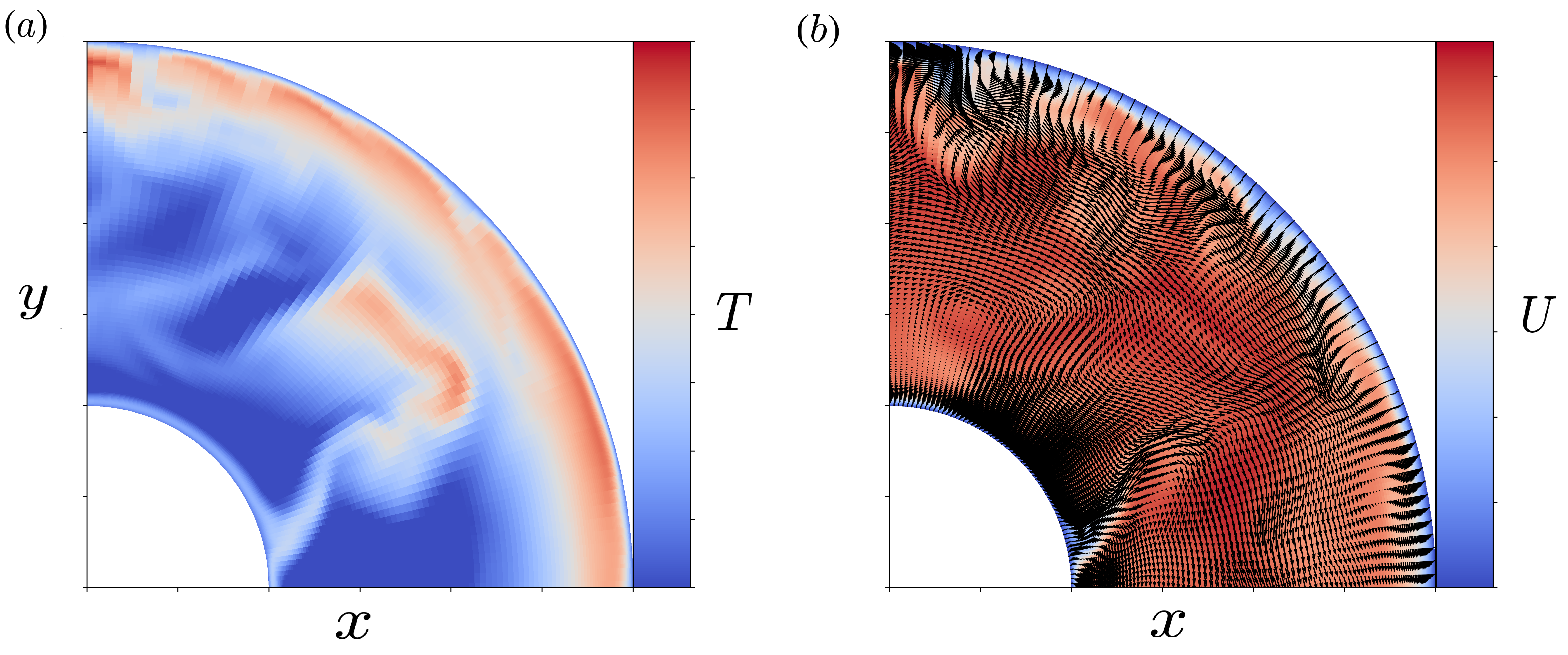}
	\caption{Centrifugal force effect on the (a) temperature field; (b) velocity field}
	\label{fig:pipe-structure}
\end{figure}

\begin{figure}
	\centering
	\includegraphics[width=\linewidth]{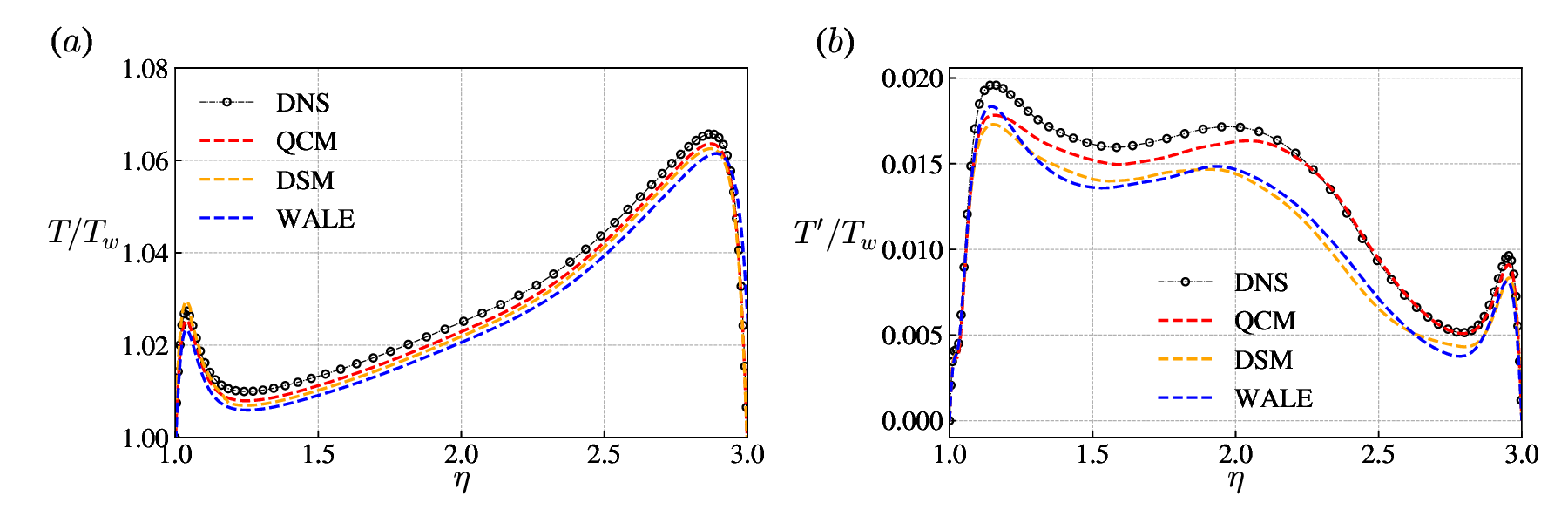}
	\caption{Wall-normal profiles of mean and fluctuating temperature: (a) mean temperature normalized by wall temperature; (b) r.m.s of temperature normalized by wall temperature.}
	\label{fig:pipe-mean-fluc-T}
\end{figure}

Figure \ref{fig:pipe-U-dns} shows the instantaneous velocity distribution in a slice from DNS results. One can notice that due to centrifugal force, flow around the inner wall is thrown to the outer wall, thus producing a strong compression effect. Fluid is heated in the near outer wall region and its density gets larger due to the compression effect, thus producing a high-pressure gradient to confront the centrifugal force, as can be seen in Figure \ref{fig:pipe-structure}. This phenomenon makes the distribution of mean and fluctuating temperature a vitally important topic. Figure \ref{fig:pipe-mean-fluc-T} shows a comparison of mean and fluctuating temperature profiles between DNS and different LES models. We can see that the result from QCM precisely accords with the real mean temperature. The DSM and the WALE model can also supply passable prediction results. As for the r.m.s of fluctuating temperature, the other two models failed to give accurate predictions in the outer wall region, while the QCM successfully gave accurate predictions.

\begin{figure}
	\centering
	\includegraphics[width=0.7\linewidth]{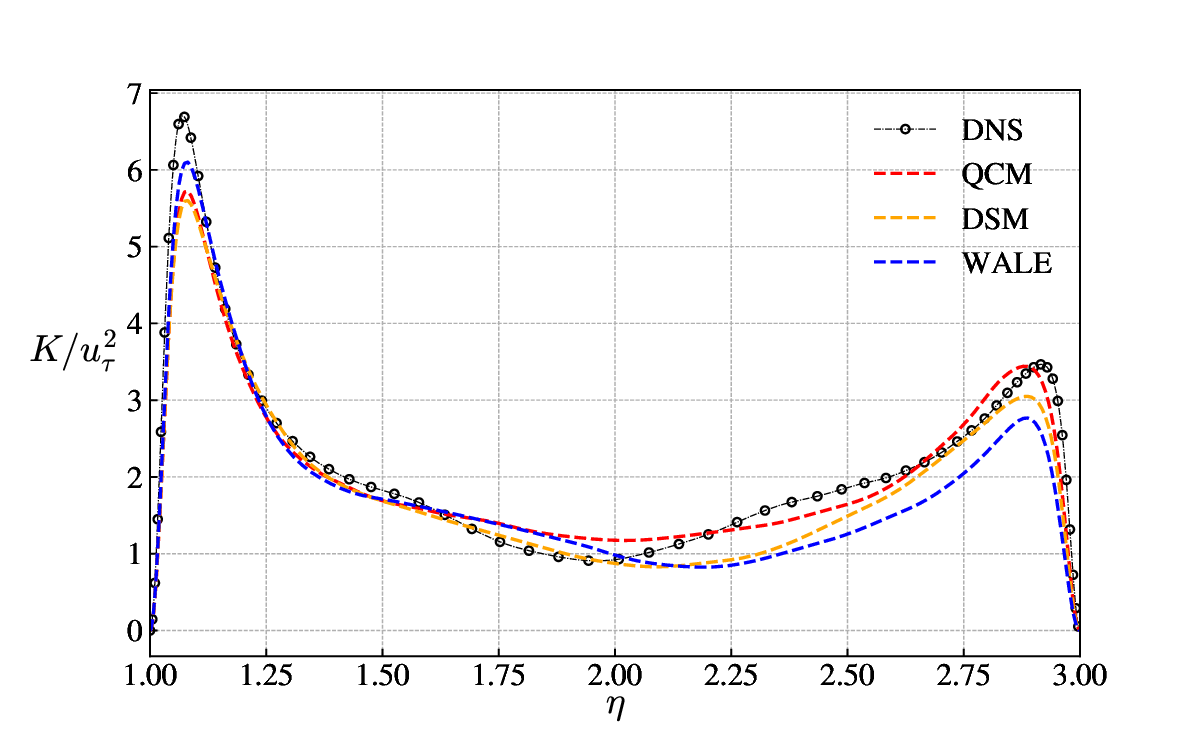}
	\caption{Distribution of normalized turbulent kinetic energy}
	\label{fig:pipe-fluc}
\end{figure}

Figure \ref{fig:pipe-fluc} shows the normalized turbulent kinetic energy distribution. All three models give better predictions in the near inner-wall region than in the near outer-wall region. The QCM shows higher kinetic energy in the near outer wall region than the other two models. In the middle part of the pipe, the QCM gives more accurate turbulent kinetic energy, while the WALE model underpredicted the turbulent kinetic energy and is seriously deviated from the DNS results.

% Figure \ref{fig:pipe-pih1} shows distribution of the first channel of helicity flux. It can be observed that all the three models gives similar distribution of helicity flux in the near inner wall region, while in the near outer wall region, the WALE model is lack of details. The DSM shows extrusion distribution in the near outer wall region. In comparison to DNS, the QCM gives a generally more similar distribution.
% \begin{figure}
% 	\centering
% 	\includegraphics[width=0.9\linewidth]{figs/pih1.pdf}
% 	\caption{Distribution of first channel of helicity flux, (a) DNS; (b) QCM; (c) DSM; (d) the WALE model}
% 	\label{fig:pipe-pih1}
% \end{figure}

\begin{figure}
	\centering
	\includegraphics[width=\linewidth]{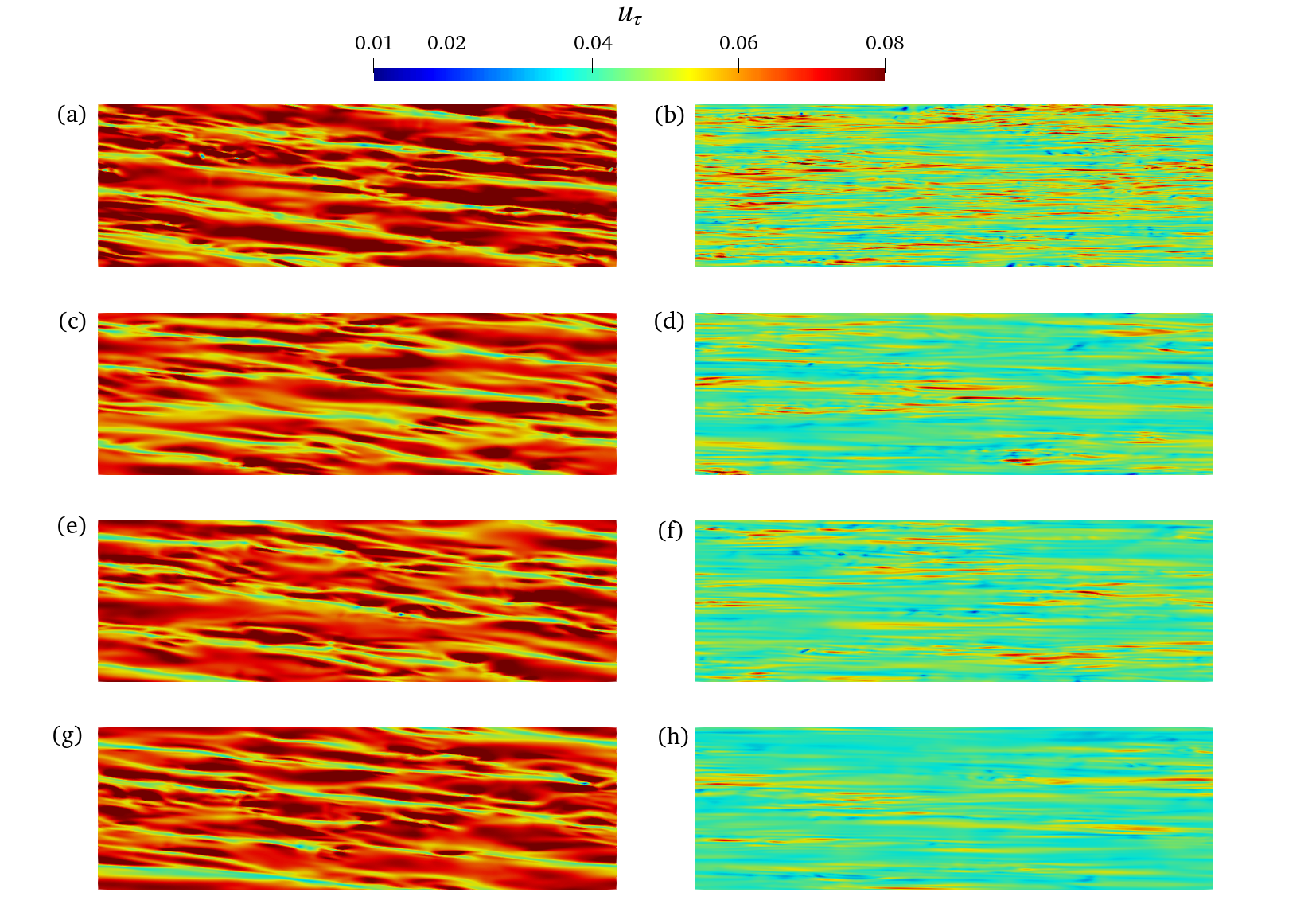}
	\caption{Friction velocity distribution on the inner and outer wall of the annular pipe flow with streamwise rotation. Left and right represent the inner and outer walls, respectively. (a, b) DNS; (c, d) the QCM; (e, f) the DSM; (g, h) the WALE model}
	\label{fig:pipe-utau}
\end{figure}

Figure \ref{fig:pipe-utau} depicts the friction velocity distribution on the inner and outer walls. On the inner wall, a large-scale strike with an inclination angle can be observed, while on the outer wall exists smaller strikes, and they are almost parallel with the flow direction. Compared with the DNS data, we can see all three models give a similar picture on the inner wall, and they can show most of the details. On the outer wall, where flow withstands larger centrifugal force, the QCM and the DSM give more abundant details than the WALE model.
% The coherent structure of the compression annular pipe flow with streamwise rotation is shown in Figure \ref{fig:pip-Q}. We can see that in the near wall region, all three models show a large scale strip structure, while the QCM can predict more abundant coherent structures, especially in the near-wall region. We have to note that according to the Q isosurface, even if the mesh is much coaser that the DNS mesh and so much information of small scale vortices are lost, the QCM model can still give rather accurate prediction of turbulence quantity distributions. This is encouraging that it shows the new model is improved in representsing the effect of small to medium scale vortices.
% \begin{figure}
% 	\centering
% 	\includegraphics[width=0.7\linewidth]{figs/dns.eps}
% 	\caption{Schemetic diagram of streamwise-rotating annular pipe}
% 	\label{fig:pip-Q}
% \end{figure}

\subsection {Application in hypersonic flow over a rotating cone}
\begin{figure}
	\centering
	\includegraphics[width=0.8\linewidth]{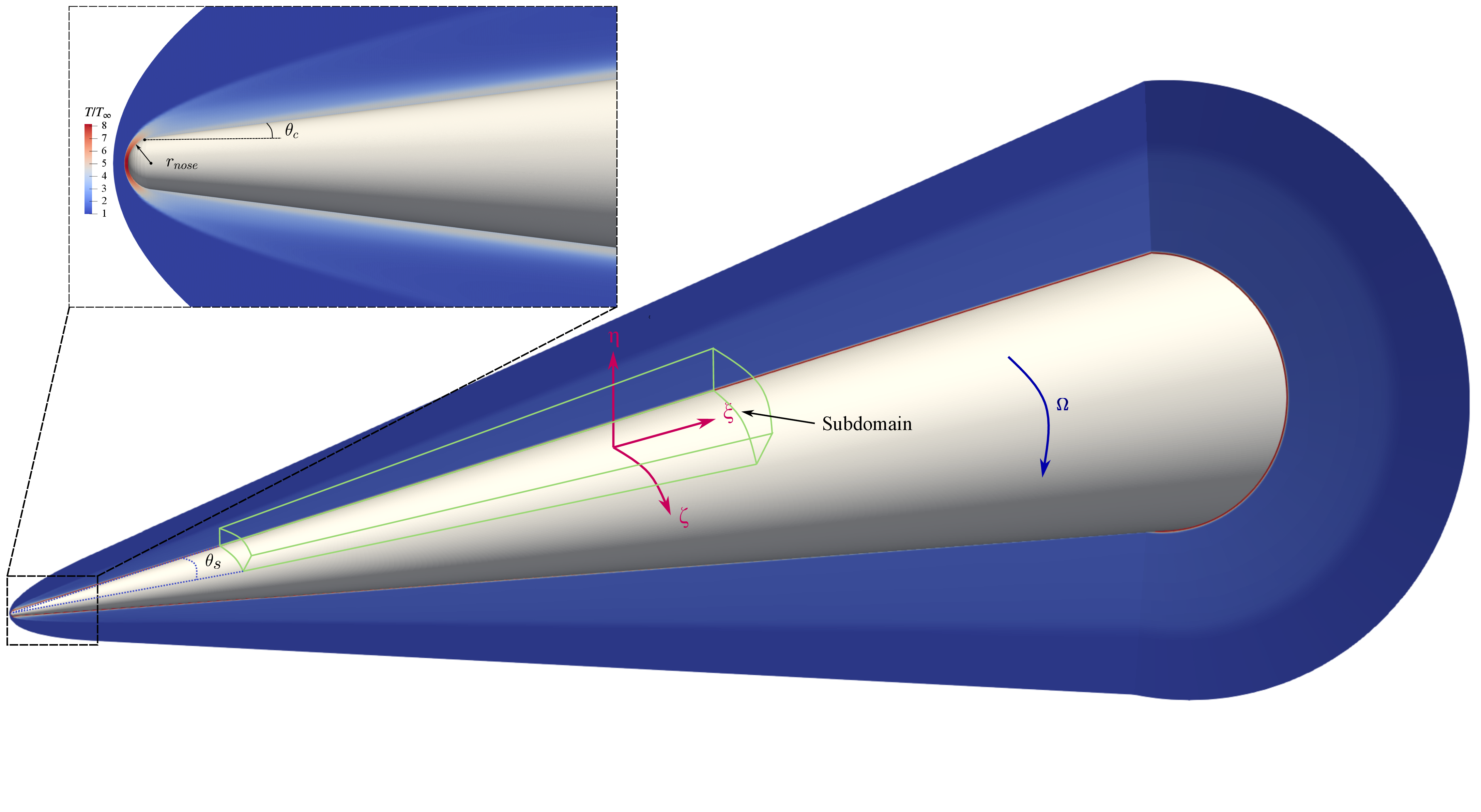}
	\caption{Schematic diagram of hypersonic flow over a streamwise-rotating blunt cone}
	\label{fig:cone-schemetic}
\end{figure}

In this section, we will test the proposed model in hypersonic ($Ma = 6$) flow over a rotation cone, to test the model's ability in predicting the transition process. The schematic diagram of the case is shown in figure \ref{fig:cone-schemetic}. The geometry is a $7^{\circ}$ half-angle straight cone with a nose radius of $r_{nose} = 1\rm{mm}$. The computation is nondimensionalized by using inflow parameters $u_{\infty}, \rho_{\infty}, T_{\infty}$, and the nose radius. Cone rotation angular velocity $\Omega$ is nondimensionalized by inflow velocity and nose radius. The flow conditions are summarized in table \ref{tab:cond-cone}, and the fluid is considered to be perfect gas with a constant Prandtl number $Pr=0.7$ and a constant ratio of specific heat ($\gamma=1.4$), and the viscosity is calculated using the standard Sutherland's law.

\begin{table}
	\begin{center}
  \def~{\hphantom{0}}
	\begin{tabular}{lccccc}
		  Ma&  $Re (\rm{m^{-1}})$ & $T_{\infty} (\rm{K})$ & $T_w (\rm{K})$ & $\theta_c (deg)$ &$\Omega$ \\[3pt]
		  $6.0$ & $6\times 10^6$ & $79$ & $300$ & $7$ & $0.002$
	\end{tabular}
	\caption{Flow conditions of hypersonic flow over rotation cone.}
	\label{tab:cond-cone}
	\end{center}
\end{table}

For the simulation presented here, a two-step strategy was employed. In the first step, steady flow is computed with the computational domain covering the cone's head and the bow shock using a finite-volume code developed by the authors. In the second step (transitional flow simulation), blow and suck disturbances are introduced in the front of the computational subdomain to trigger the Bypass transition. The subdomain is denoted in figure \ref{fig:cone-schemetic}. By using high-order finite-difference code, the transition process can be accurately captured. Like the numerical settings of compressible annular pipe flow case, in DNS simulation, we adopt a mixed scheme for convective discretization and an eighth-order central scheme for viscous term discretization. In LES, a six-order central difference scheme is used for convective term discretization. To ensure numerical stability, a low dissipation filter is employed to eliminate non-physical high-frequency oscillation for the LES simulations. Time integration is achieved by the third-order Runge-Kutta method. Computational mesh settings of unsteady (transition) simulation are shown in table \ref{tab:cone-grid}. The computational subdomain is located inside the bow shock, the boundary condition for the upper boundary of the subdomain is the Dirichlet boundary, and the value is obtained from steady computation interpolation. Periodic boundary conditions are applied at the Circumferential boundaries. To remove numerical disturbances reflected from the outlet boundary, mesh coarsening toward the outlet boundary is employed in this study. 

\begin{table}
	\begin{center}
  \def~{\hphantom{0}}
	\begin{tabular}{lccccc}
		  Case& $N_{\xi}\times N_{\eta}\times N_{\zeta}$   &  $\Delta \xi^{+}$ & $\Delta \eta_{min}^{+}$ & $\Delta \eta_{max}^{+}$ & $\Delta \zeta_{max}^{+}$\\[3pt]
		 DNS   & $5000\times 250 \times 500$ & 8.48 & 0.54& 26.46 & 15.0\\
		 LES   & $1800 \times 100 \times 200$ & 24.3 & 0.54 & 54.0& 37.5\\
	\end{tabular}
	\caption{Grid settings and grid resolutions of the simulations in the hypersonic flow over a streamwise rotating cone.}
	\label{tab:cone-grid}
	\end{center}
\end{table}

\begin{figure}
	\centering
	\includegraphics[width=0.9\linewidth]{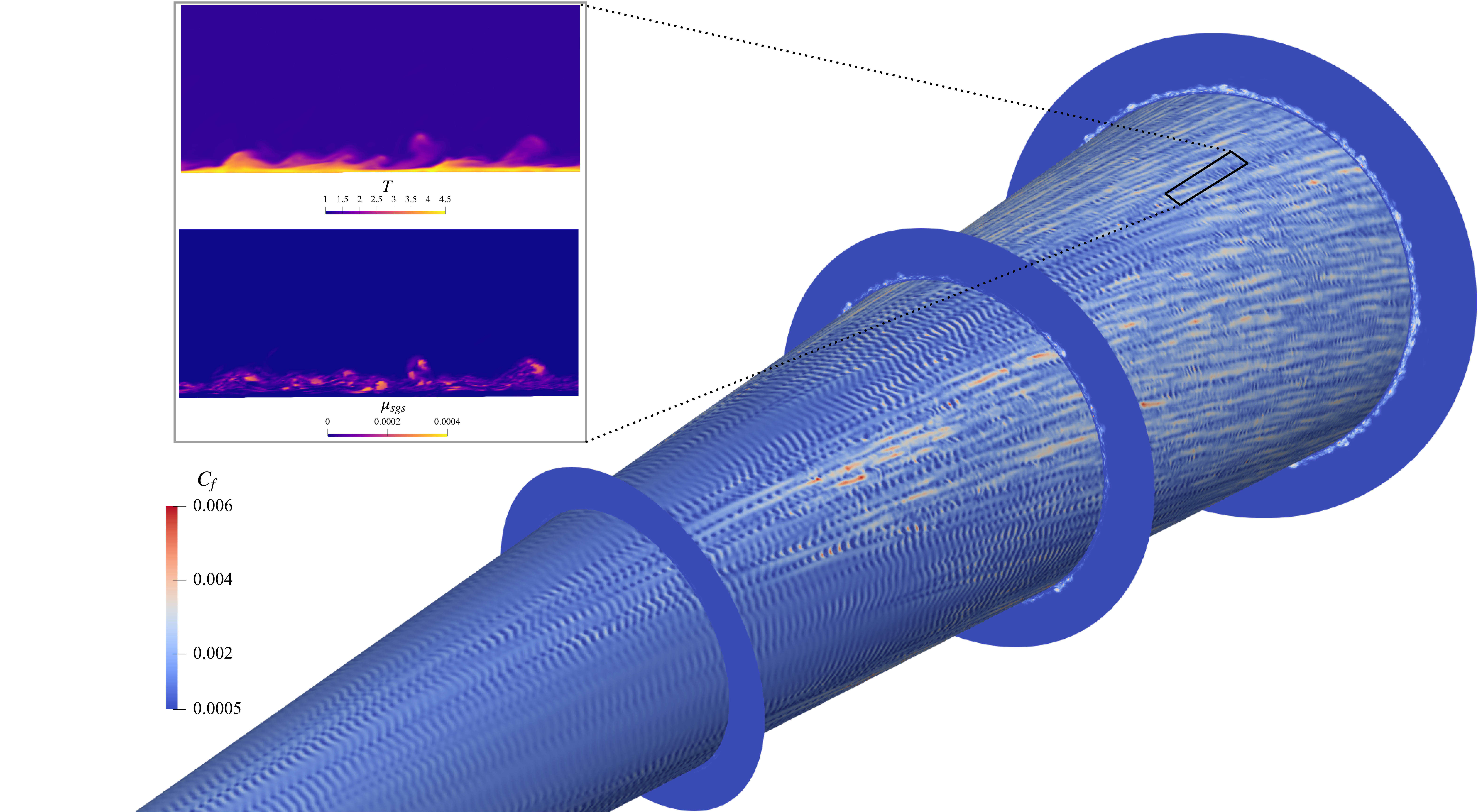}
	\caption{Distribution of SGS viscosity}
	\label{fig:cone-mut}
\end{figure}

Figure \ref{fig:cone-mut} shows the distribution of SGS viscosity $\mu_{sgs}$ along the cone. Since the SGS kinetic energy is very small in the pre-transitional flow, the QCM model provides nearly zero eddy viscosity there, allowing small-amplitude disturbances to grow in the pre-transitional flow. Once small eddies are formed, the intrinsic dependency of helicity on the vorticity triggers the rapid growth of the model coefficients and eventually the eddy viscosity in the late stage of the transition and the subsequent turbulent region. By introducing the joint constraint, the new model considers the effect of helicity, which may provide proper distribution of eddy viscosity. 

\begin{figure}
	\centering
	\includegraphics[width=\linewidth]{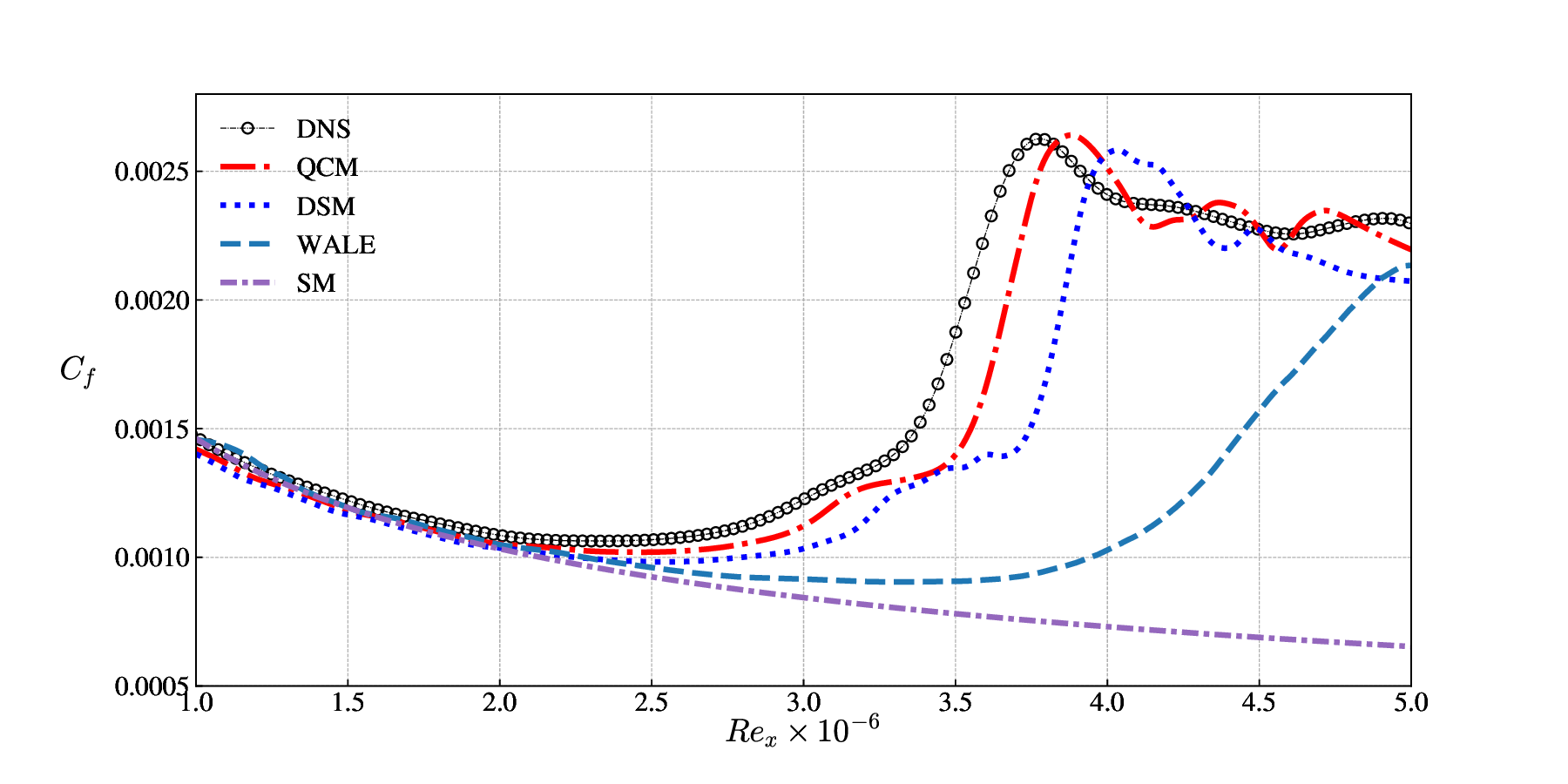}
	\caption{Distribution of skin friction coefficient as a function of streamwise Reynolds number}
	\label{fig:cone-cf}
\end{figure}

Figure \ref{fig:cone-cf} shows the profile of skin friction distribution in the streamwise direction. We can see that the transition process is extremely sensitive to the SGS viscosity provided by the LES model. The constant-Smagorinsky model is too dissipative and fails to provide a sufficiently small eddy viscosity in the pre-transition region, causing the given disturbances to dampen out and preventing the flow from undergoing turbulent transition in the entire computational domain. The WALE model also fails to reach fully turbulent during the whole process. The DSM and the QCM give similar transition start points and the result from QCM is much better than DSM. We can get the conclusion that the QCM has good performance in predicting transitional flow, which is reflected in two ways: reduce significantly the eddy viscosity in the pre-transition region and provide enough accurate eddy viscosity in the fully turbulent region.

\section{Conclusions}
A quasi-dynamic one-equation model with joint constraints of kinetic energy and helicity fluxes (QCM) is proposed for large-eddy simulation (LES) of both incompressible and compressible flows in this paper. We start from the deconvolution model, the subgrid-scale (SGS) kinetic energy equation is also introduced to constrain the model coefficient. Through this process, a more accurate SGS stress can be obtained. With the precise SGS stress, an accurate dual-channel helicity flux can be obtained, which is vitally important for rotating turbulent flows that have large-scale helical motion and energy backscatter. To improve the numerical stability of the deconvolution model, dual channels of helicity flux, along with kinetic energy flux are used to constrain the Smagorinksy model. A weighted joint constraint is adopted to ensure the model coefficient is within a reasonable range, which helps maintain numerical stability. During the quasi-dynamic process, all the coefficients of modelled quantities are resolved dynamically without test filtering, which is especially beneficial for LES of complex geometry and Highly distorted anisotropic mesh.
At the \textit{a priori} test of incompressible turbulent channel flow with streamwise rotation, the new model shows a high correlation with the real SGS stress, and it also shows that the SGS kinetic energy, kinetic energy flux, and helicity flux maintain a high correlation both in the near-wall region and in the reverse flow region near the channel centre.

The new model is first examined in incompressible channel flow with streamwise rotation at two different Reynolds numbers. The QCM model can well predict turbulent quantities such as mean velocity, turbulence intensities, mean helicity, fluctuating helicity and energy, etc. Compared to DSM and WALE models, the QCM shows improved predicting ability in both the near-wall region where a secondary flow exists and the reverse flow region near the channel centreline. The new model is proved to be better than the other two models at a higher Reynolds number. For the case of compressible annular pipe flow with streamwise rotation, the QCM model can accurately predict quantities related to compressible turbulent flow, which include the mean temperature profile, the turbulent kinetic energy, and the friction velocity distribution. The new model also shows the capability of capturing the elaborated structure of turbulence. For the last case of transitional flow over a rotating hypersonic cone, the QCM can well predict the transition process and gives turbulence onset position more accurately than the other models. The new model also provides a more precise skin friction coefficient compared to the other two models.

To summarize, the appearance of the Coriolis force and centrifugal force has a significant impact on the turbulent flows, by introducing the joint constraints of kinetic energy and helicity fluxes, we developed a new LES model that can depict the backscatter process while maintaining numerical stability. The newly proposed QCM is a quasi-dynamic one-equation model in which coefficients are obtained dynamically without test filtering. It is based on a procedure that tries to solve SGS kinetic energy accurately so that the QCM can depict the turbulent cascades accurately. With the new joint constraint, the new model has the ability to accurately predict complex rotating flows.

\backsection[Acknowledgements]{The authors thank the National Supercomputer Center in Tianjin (NSCC-TJ) and the National Supercomputer Center in GuangZhou (NSCC-GZ) for providing computer time.}

\backsection[Funding]{This work was supported by the National Key Research and Development Program of China (grant nos 2020YFA0711800 and 2019YFA0405302) and NSFC Projects (nos 12072349, 12232018, 12202457), National Numerical Windtunnel Project, Science Challenge Project (grant no. TZ2016001) and Strategic Priority Research Program of Chinese Academy of Sciences (grant no. XDC01000000).}

\backsection[Declaration of interests]{The authors report no conflict of interest.}

\backsection[Author ORCIDs]{Changping Yu, https://orcid.org/0000-0002-2126-1344}

\appendix

\bibliographystyle{jfm}
\bibliography{jfm}

\end{document}